\documentclass[11pt]{article}
\usepackage{amsmath, amssymb}
\usepackage{slashed}
\usepackage{verbatim}
\usepackage{latexsym}
\usepackage{euscript}
\usepackage{amsfonts}
\usepackage{verbatim}
\usepackage[]{array}
\usepackage[]{mathrsfs}
\usepackage{graphicx}
\usepackage{amsmath}
\usepackage{amssymb}
\usepackage{amsxtra}
\def\be{\begin{eqnarray}}
\def\ee{\end{eqnarray}}
\def\0{\nonumber}

\def\tr{{\rm tr}}
\def\Tr{{\rm Tr}}

\usepackage{slashed}
\usepackage{verbatim}
\usepackage{latexsym}
\usepackage{euscript}
\newcommand\ET{\EuScript{T}}
\newcommand\EG{\EuScript{G}}
\newcommand\EF{\EuScript{F}}

\newcommand\ES{\EuScript{S}}

\normalfont\large
\begin{document}
\begin{flushright}
SISSA/26/2018/FISI\\{ZTF-EP-18-02}\\
arXiv:1807.01249
\end{flushright}
\vskip 2cm
\begin{center}

{\LARGE {\bf Axial gravity: a non-perturbative approach to split anomalies}}

\vskip 1cm

{\large  L.~Bonora$^{a}$, M.~Cvitan$^{b}$, P.~Dominis
Prester$^{c}$, S.~Giaccari$^{d}$,
M.~Pauli\v{s}i\'c$^{c}$, T.~$\rm \check{S}$temberga$^{b}$
\\\textit{${}^{a}$ International School for Advanced Studies (SISSA),\\Via
Bonomea 265, 34136 Trieste, Italy, and INFN, Sezione di
Trieste \\ }
\textit{${}^{b}$ Department of Physics, Faculty of Science, University of
Zagreb, Bijeni\v{c}ka cesta 32,
10000 Zagreb, Croatia \\}
\textit{${}^{c}$ Department of Physics, University of Rijeka,\\
Radmile Matej\v{c}i\'{c} 2, 51000 Rijeka, Croatia\\}
\textit{${}^{d}$ Department of Sciences,
Holon Institute of Technology (HIT),\\
52 Golomb St., Holon 5810201, Israel}
}
\vskip 1cm

\end{center}

 \vskip 1cm {\bf Abstract.}
In a theory of a Dirac fermion field coupled to a metric-axial-tensor (MAT)
background, using
a Schwinger-DeWitt heat kernel technique,  
we compute non-perturbatively the two (odd parity) trace anomalies. 
A suitable collapsing limit of this model 
corresponds to a theory of chiral fermions coupled to (ordinary) gravity. Taking
this limit 
on the two computed trace anomalies we verify that they tend to the same
expression, 
which coincides with the already found odd parity trace anomaly, with the
identical
coefficient. 
This confirms our previous results on this issue.
\vskip 1cm
\begin{center}

{\tt Email: bonora@sissa.it,mcvitan@phy.hr,
pprester@phy.uniri.hr,stefanog@hit.ac.il,mateo.paulisic@phy.uniri.hr,
tstember@phy.hr}

\end{center}

\eject

\section{Introduction}\label{s:intro}

This paper is a follow up of \cite{MAT1} where a new version of modified gravity
was introduced, a 
metric-axial-tensor gravity. That is, beside the usual metric, the model is
endowed with an additional 
symmetric tensor that interacts chirally with fermions. The purpose there was
not (or not yet) to 
describe a new phenomenological model of gravity, but to permit a more accurate
investigation 
of the relation between gravity and chiral fermions. It is often stated in the
literature that 
gravity is chirally blind, meaning that the relevant charge, the mass, is
positive, and is 
thus different from the typical case of a U(1) interaction. This is certainly a
basic peculiarity 
of gravity with several important consequences. However one should reflect on
the fact that the coupling
between gravity and matter is given by the juxtaposition of the metric and the
energy-momentum tensor, 
and the energy-momentum tensors of fermions with opposite chiralities are
different. 

 {One can suspect} therefore that at some stage 
differences  {might} emerge between fermions with opposite chiralities in their
interaction with gravity.
A privileged place where such differences may show up are the anomalies. And in
this case the 
candidate is the trace anomaly, because it involves precisely the coupling
between the metric 
and the energy-momentum tensor. The difficulty is how to make this difference to
emerge. As will be 
argued below, one should be careful to preserve the definite fermion chirality
throughout the 
calculation. There is no direct way to do it, basically because the Dirac
operator for a Weyl 
fermion contains a chiral projector. Therefore one has to resort to some
indirect method. Like in many other cases in physics, the best way to avoid
similar problems is to embed the system  in a larger setup containing more
variables and/or parameters.  The metric-axial-tensor (MAT) gravity is
designed to do this. It is formulated for Dirac fermions 
coupled to the usual metric and to an axial symmetric tensor. In this case the
operator involved is the usual 
Dirac operator. The situation appropriate for Weyl fermions is recovered in a
specific limit, the collapsing limit.

As mentioned above, MAT has already been introduced and used to compute the
odd-parity trace anomaly in \cite{MAT1}. 
There the approach was perturbative, we calculated the Feynman diagrams at the
lowest significant order. What we want to do in this paper is to show that the
same result can be obtained non-perturbatively, by means of the heat kernel
method and using different regularizations. Hereafter is a qualitative, but more
detailed, presentation of both the problem we wish to solve and the method we
use.

\subsection{Split and non-split anomalies}

A basic differentiation between anomalies in fermionic field theories is the
separation between {\it split} and {\it non-split} anomalies. Split anomalies
have an opposite sign for opposite fermion chiralities. Non-split anomalies have
the same sign for opposite chiralities. An example of the first are the
consistent chiral gauge or gravity anomalies. They may of course arise only in
the 
presence of chiral asymmetry. These anomalies undermine the consistency of
theories in which they are present, and, as a consequence, they have been used
as an exclusion criterion.  An example of non-split anomalies are the covariant
gauge or gravity anomalies, such as the
Kimura-Delbourgo-Salam anomaly or the anomaly  that is utilized to explain the
decay of a $\pi^0$ into two $\gamma$'s. But the examples are manifold. In the
family of trace anomalies, the even ones are non-split, while the odd trace
anomaly,
which is the main character of this paper, is split.

Split and non-split anomalies differ also for the difficulties one comes across
when computing them. While there are several tested techniques to compute
non-split anomalies, the calculation of the split ones is rather non-trivial. In
many of the latter cases one may avail oneself of such a powerful tool as the
family index theorem (for instance for consistent gauge and gravity anomalies).
But, like for the odd trace anomaly, this is not always so, and, in any case, it
is important to be able to derive such anomalies with independent
field-theoretical methods. If one resorts to path integral methods, one has to
integrate out the fermion field(s), in which case the origin of the difficulties
resides in the functional measure.
As discussed in \cite{MAT1}, a basic ingredient for the calculation is the
functional integration measure which, for chiral fermions, is not well-defined.
On the other hand, to get the correct result, it is imperative to preserve
throughout the calculation the information that the fermion field, which is
being integrated out, has a definite chirality. One is then obliged to either
use indirect methods or to elude a direct intrusion of the functional measure in
the calculation. The second alternative refers to the use of Feynman diagrams,
in which case the chirality of fermions is preserved by vertices containing the
appropriate chiral projector. This is the method employed in
\cite{BGL,BDL,MAT1}
together with dimensional regularization. In the present paper however, we focus
on an indirect method of calculation, first used by Bardeen, \cite{Bardeen}, for
chiral gauge anomalies. He considered a theory of Dirac fermions coupled
to two external non-Abelian (vector $V_\mu$ and axial $A_\mu$) gauge
potentials.
Clearly this poses no problems from the point of view of the functional measure
and
the derivation of the anomaly goes through without difficulties. Eventually
one takes the collapsing limit $V\to \frac V2$ and $A\to \frac V2$ and verifies
that,
in such a limit,
the anomaly becomes the desired consistent gauge anomaly. For the sake of
clarity we
present a summary of this derivation in Appendix \ref{s:bardeen}.

This approach has already been introduced and applied in \cite{MAT1} for the
odd trace anomaly.
To this end we introduced there a modification of ordinary gravity,
the metric-axial-tensor (MAT) gravity:
beside the usual metric $g_{\mu\nu}$ we introduced an axial symmetric 2-tensor
$f_{\mu\nu}$,
and coupled it to a Dirac fermion. Then we computed the trace of the
energy-momentum tensor
and of its axial companion and, eventually, we took the limit $g\to \frac g2$
and $f\to \frac g2$ and obtained the desired result.
The limit of that derivation is that
it relies on Feynman diagram techniques, and, so, it is perturbative. In fact we
calculated
only the lowest order of the odd trace anomaly and then covariantized it. This
is of course
permitted provided we are sure that there are no anomalies of the
diffeomorphisms.
With a MAT background this verification is exceedingly complicated and in
\cite{MAT1} we
did not do it and contented ourselves with an analogous but simpler verification
carried
out in \cite{BDL}. It is clear that to prevent any objection we have to
guarantee that
diffeomorphisms are respected throughout the derivation. This can be done with
DeWitt's method,
\cite{DeWitt1,DeWitt2}. This method is based on point-splitting. Therefore one
needs a regularization
in order to get rid of divergences, but the point-splitting is along a geodesic,
thus guaranteeing covariance under diffeomorphisms. Our aim here is to combine
DeWitt's with Bardeen's method.
This requires a formulation of MAT more accurate than in \cite{MAT1}. For this
reason the anomaly calculation proper needs to be preceded by a long
introduction on the
so-called hypercomplex calculus, which is the appropriate framework for MAT
gravity.

\vskip 0.5 cm
{\bf Organization of the paper}. Section 2 is a short introduction of
axial-complex numbers and axial-complex analysis. In section 3 we
deal with the axial-complex analysis of geodesics in an axial-complex space. We
introduce normal coordinates, define the world function and the coincidence
limit (i.e. the limit for vanishing geodetic distance), the VVM determinant and
the parallel displacement matrix for tensors and for spinors. The
(pseudo)Riemannian geometry of an axial-complex space was already introduced in
\cite{MAT1}. To help the reader, it is presented anew in Appendix B in a
partially renovated notation, which seems to us more practical.  In section 4 we
introduce the theory of Dirac fermions in a MAT background, we define the
relevant energy-momentum tensors (they are two, the ordinary one and its axial
companion)  and analyse their classical Ward identities with respect to ordinary
and axial diffeormorphisms and Weyl transformations. We also define the `square'
of the Dirac operator, which is crucial for the application of the
Schwinger-DeWitt method. In section 5 we explain this method and compute the
relevant heat kernel coefficients. In section 6 we apply these results to the
non-perturbative computation of the (odd) trace anomalies of the two em tensors
with two different regularization, the dimensional and $\zeta$-function ones.
Then we compute the collapsing limit and show that the two anomalies collapse to
a single one and take the form of the odd trace anomaly already computed in 
\cite{BGL,BDL} and \cite{MAT1}, as expected. Section 7 is devoted to our
conclusions. Appendix A is a summary of Bardeen's method. Appendix C contains a
short account of fermion propagators in a  MAT background.

{
 \vskip 0.5 cm
{\bf Overview of the literature.} There exists a vast literature on even trace
anomalies in 4d, mostly old [9-29]  but also recent
\cite{Bzowski,Godazgar,Coriano}, 
denoting a renewed interest in the subject. The literature
on the odd parity trace anomaly in 4d (still in a settling phase) consists of
\cite{MAT1,BGL,BDL,BL,Shapiro,Bast,Nakayama1,Nakayama2}.
Textbooks on anomalies are \cite{Bertlmann,Fujikawa,BastVan}. Aspects of split
and non-split anomalies, 
which are relevant to this paper, were discussed in \cite{AB1,AB2}. A
regularization, not used in 
this paper, but which would be interesting to explore is the one introduced in
\cite{Grabowska}.
Hypercomplex analysis in physical problems was introduced and used in 
\cite{Crumey,Clerc,Mantz,Hess-Greiner}.
}

\section{Axial-complex analysis}

Axial-complex numbers are defined by
\be
\hat a = a_1+\gamma_5 a_2\label{hata}
\ee
where $a_1$ and $a_2$ are real numbers. Arithmetic is defined in the obvious
way. We can
define a conjugation operator
\be
\overline {\hat a} = a_1-\gamma_5 a_2\label{conjugation}
\ee
We will denote by ${\cal A}{\cal C}$ the set axial-complex numbers, by ${\cal
A}{\cal R}$ the set of axial-complex numbers with $a_2=0$
(the axial-real numbers) and by
${\cal A}{\cal I}$ the set of axial-complex numbers with $a_1=0$ (the
axial-imaginary numbers).
We can define a (pseudo)norm
\be
(a,a)= \hat a\overline {\hat a} = a_1^2-a_2^2\label{pseudonorm}
\ee
This determines an axial-light-cone with all the related problems. In general,
whenever possible, we will keep away from it by considering the case $|a_1|
>|a_2|$. Alternatively we will use an axial-Wick-rotation (analogous to the Wick
rotation for the Minkowski spacetime light-cone) $a_2 \to i a_2$. Whenever we
resort to it explicit mention will be made.

Introducing the chiral projectors $P_\pm=\frac {1\pm\gamma_5}2$, we can also
write
\be
\hat a= a_+P_++ a_- P_-,\quad\quad a_\pm = a_1\pm a_2\label{a+-}
\ee
We will consider functions $\hat f(\hat x)$ of the axial-complex variable
\be
\widehat x= x_1+ \gamma_5 x_2 \label{xhat}
\ee
from ${\cal A}{\cal C}$ to  ${\cal A}{\cal
C}$, which are axial-analytic, i.e. admit a Taylor expansion, and
actually identify the functions with their expansions. Using the property of the
projectors it is easy to see that
\be
\hat f(\hat x) = P_+ { \hat f}(x_+)+  P_- { \hat f}(x_-) =\frac 12 \left(\hat
f(x_+)+\hat
f(x_-)\right) + \frac {\gamma_5}2 \left(\hat f(x_+)-\hat
f(x_-)\right)\label{fhatx}
\ee
In the same way we will consider functions from ${\cal A}{\cal C}^4$ to ${\cal
A}{\cal C}$, with analogous properties.
\be
\hat f(\hat x^\mu) = P_+ {\hat f}(x^\mu_+)+  P_- {\hat f}(x^\mu_-) =\frac 12
\left(\hat
f(x^\mu_+)+\hat f(x^\mu_-)\right)
+ \frac {\gamma_5}2 \left(\hat f(x^\mu_+)-\hat f(x^\mu_-)\right)\label{fhatx4}
\ee
with $\mu=0,1,2,3$, and
\be
\widehat x^\mu= x_1^\mu + \gamma_5 x_2^\mu \label{xmuhat}
\ee
are the axial-complex coordinates.

Axial-complex numbers and analysis are a particular case of pseudo-complex or
hyper-complex numbers and analysis, \cite{Hess-Greiner}.

Derivatives are defined in the obvious way:
\be
\frac {\partial}{\partial \hat x^\mu} = \frac 12 \left( \frac
{\partial}{\partial  x_1^\mu}+\gamma_5 \frac {\partial}{\partial
x_2^\mu}\right),
\quad\quad
\frac {\partial}{\partial {\overline{\hat x}} ^\mu}
= \frac 12 \left( \frac {\partial}{\partial  x_1^\mu}-\gamma_5 \frac
{\partial}{\partial  x_2^\mu}\right)
\ee
Notice that for axial-analytic functions
\be
\frac d{d\hat x} = {\frac \partial{\partial x_1}\equiv\frac {\partial}{\partial
\hat x},\label{dxdx1}}
\ee
{whereas $\frac {\partial}{\partial {\overline{\hat x}}}\widehat f(\hat x)=0$.}

As for integrals, since we will always have to do with rapidly decreasing
functions at infinity, we define
\be
\int  d\hat x\,\widehat f(\hat x)\0
\ee
as the rapidly decreasing primitive $\widehat g(\hat x)$ of $\widehat f(\hat
x)$. Therefore the property
\be
\int d\hat x\, \frac {\partial}{\partial\hat x^\mu} \hat
f(\hat x)=0\label{stokes}
\ee
follows immediately. As a consequence of \eqref{dxdx1} it follows that, for an
axial-{analytic} function,
\be
\int  d\hat x\,\widehat f(\hat x)= \int  d x_1\,\widehat f(\hat
x)\label{dxdx1int}
\ee
and we can define definite integrals such as
\be
\int_{\hat a}^{\hat b} d\hat x \, \widehat f(\hat x) = \widehat g(\hat
b)-\widehat g(\hat a)\label{defint}
\ee

In this axial-spacetime we introduce an axial-Riemannian geometry as follows.
Starting from a metric $\widehat g_{\mu\nu} = g_{\mu\nu}+\gamma_5 f_{\mu\nu}$,
the
Christoffel symbols (see Appendix {B}) are defined by

\be
\widehat \Gamma_{\mu\nu}^\lambda &=& \frac 12 \widehat
g^{\lambda\rho}\left(
\frac{\partial}{
\partial{\widehat x^\mu}} \widehat g_{\rho\nu}
+
\frac{\partial}{ 
\partial{\widehat x^\nu}} \widehat g_{\mu\rho}
- 
\frac{\partial}{
\partial{\widehat x^\rho}}
\widehat
g_{\mu\nu}\right)\label{Christ}%
\ee
They split as follows
\be
\widehat\Gamma_{\nu\lambda}^\mu  = \Gamma_{\nu\lambda}^{(1)\mu}  +\gamma_5
\Gamma_{\nu\lambda}^{(2)\mu}\label{Gammahat}
\ee
and are such that the metricity condition is satisfied
\be
\frac{\partial}{\partial \hat x^\mu} \widehat g_{\nu\lambda} =  \widehat
\Gamma_{\mu\nu}^\rho\, \widehat g_{\rho\lambda} +
 \widehat \Gamma_{\mu\lambda}^\rho \,\widehat g_{\nu\rho},\label{metricity1}
\ee
which, in ${\cal A}{\cal R}^4$, takes the form
\be
&&\frac{\partial}{\partial {\hat x}^\mu} g_{\nu\lambda} =
\Gamma_{\mu\nu}^{(1)\rho}\, g_{\rho\lambda} +
\Gamma_{\mu\lambda}^{(1)\rho} \, g_{\nu\rho}+  \Gamma_{\mu\nu}^{(2)\rho}\,
f_{\rho\lambda} +
\Gamma_{\mu\lambda}^{(2)\rho} \, f_{\nu\rho}\label{metricity2}\\
&&\frac{\partial}{\partial {\hat x}^\mu} f_{\nu\lambda} =
\Gamma_{\mu\nu}^{(1)\rho}\, f_{\rho\lambda} +
\Gamma_{\mu\lambda}^{(1)\rho} \, f_{\nu\rho}+  \Gamma_{\mu\nu}^{(2)\rho}\,
g_{\rho\lambda} +
\Gamma_{\mu\lambda}^{(2)\rho} \, g_{\nu\rho}\label{metricity3}
\ee

\section{MAT geodesics}

Let us set
\be
\widehat\Gamma_{\nu\lambda}^\mu  = \Gamma_{\nu\lambda}^{(1)\mu}  +\gamma_5
\Gamma_{\nu\lambda}^{(2)\mu}\label{Gamma}
\ee
The equation for MAT geodesics is
\be
\ddot {\widehat x}^\mu + \widehat \Gamma_{\nu\lambda}^\mu \dot {\widehat
x}^\nu\dot {\widehat
x}^\lambda=0\label{geodesic}
\ee
where a dot denotes derivation with respect to an axial-affine parameter
$t=t_1+\gamma_5 t_2$.
For axial-real and axial-imaginary components this means
\be
&&\ddot {x}_1^\mu + \Gamma_{\nu\lambda}^{(1)\mu}( \dot {x}_1^\nu\dot
{x}_1^\lambda +  \dot {x}_2^\nu\dot {x}_2^\lambda) +
\Gamma_{\nu\lambda}^{(2)\mu}( \dot {x}_1^\nu\dot {x}_2^\lambda +  \dot
{x}_2^\nu\dot {x}_1^\lambda)=0
\label{geodesic1}\\
&& \ddot {x}_2^\mu + \Gamma_{\nu\lambda}^{(1)\mu}( \dot {x}_1^\nu\dot
{x}_2^\lambda +  \dot {x}_2^\nu\dot {x}_1^\lambda) +
\Gamma_{\nu\lambda}^{(2)\mu}( \dot {x}_1^\nu\dot {x}_1^\lambda +  \dot
{x}_2^\nu\dot {x}_2^\lambda)=0
\label{geodesic2}
\ee
These geodesic equations can be obtained as equations of motion from the
action
\be
\widehat S= \int d\hat t \sqrt{\widehat g_{\mu\nu} \dot {\widehat x}^\mu\dot
{\widehat x}^\nu}=S_1+
\gamma_5 S_2 \label{hatS}
\ee
where $\widehat g_{\mu\nu} = g_{\mu\nu}+ \gamma_5 f_{\mu\nu}$.

***

The action takes values in ${\cal A}{\cal C}$. For instance, setting the
proper time $\hat \tau=\tau_1+\gamma_5\tau_2$,
\be
\widehat S[{\widehat x}] = \int d\hat\tau \left( \widehat g_{\mu\nu}
\dot{\widehat x}^\mu \dot{\widehat x}^\nu \right)^{\frac 12}
\label{Sgeo}
\ee
But unlike \cite{Hess-Greiner} we require the action principle to be specified
by $\delta \widehat S[{\widehat x}]=0$.

Taking the variation of $S[{\widehat x}]$ with respect to $\delta \widehat x =
\delta x_1 + \gamma_5 \delta x_2$, with
\be
&&\delta \widehat g_{\mu\nu} = \frac {\partial \widehat g_{\mu\nu}}{\partial
\widehat x^\lambda} \delta \widehat x^\lambda,\quad \quad {\rm i.e.}\0\\
&&{
\delta g_{\mu\nu} =  \frac12
\left(
\frac {\partial   g_{\mu\nu}}{\partial  x_1^\lambda} + 
\frac {\partial   f_{\mu\nu}}{\partial  x_2^\lambda}
\right)
\delta x_1^\lambda +
\left(
\frac {\partial   f_{\mu\nu}}{\partial  x_1^\lambda} +
\frac {\partial   g_{\mu\nu}}{\partial  x_2^\lambda} 
\right)
\delta x_2^\lambda
}\0\\
&& {\hphantom{\delta g_{\mu\nu}}=\frac {\partial   g_{\mu\nu}}{\partial 
x_1^\lambda}
\delta
x_1^\lambda +
\frac {\partial   f_{\mu\nu}}{\partial  x_1^\lambda} \delta x_2^\lambda
}\0\\
&&{
\delta f_{\mu\nu} =  \frac12
\left(
\frac {\partial   g_{\mu\nu}}{\partial  x_1^\lambda} + 
\frac {\partial   f_{\mu\nu}}{\partial  x_2^\lambda}
\right)
\delta x_2^\lambda +
\left(
\frac {\partial   f_{\mu\nu}}{\partial  x_1^\lambda} +
\frac {\partial   g_{\mu\nu}}{\partial  x_2^\lambda} 
\right)
\delta x_1^\lambda
}
\0\\
&& {\hphantom{\delta f_{\mu\nu}} =  \frac {\partial   g_{\mu\nu}}{\partial 
x_1^\lambda} \delta
x_2^\lambda +
\frac {\partial   f_{\mu\nu}}{\partial  x_1^\lambda} \delta
x_1^\lambda}
\label{deltaghat}
\ee
we get the eom
\be
{\widehat g_{\mu\rho} \ddot{\widehat x}^\rho +\widehat
\Gamma_{\nu\lambda}^\rho\,
\widehat g_{\mu\rho}\, \dot{\widehat x}^\mu \dot{\widehat x}^\nu=0},
\quad\quad {\rm i.e.} \quad\quad
\ddot {\widehat x}^\mu + \widehat \Gamma_{\nu\lambda}^\mu \dot {\widehat
x}^\nu\dot {\widehat
x}^\lambda=0\label{eomgeo}
\ee

Let us rewrite
\be
\sqrt{\widehat g_{\mu\nu} \dot {\widehat x}^\mu\dot {\widehat x}^\nu}&=&
\sqrt{A+\gamma_5 B},\label{gAB}\\
A &=& g_{\mu\nu} \left(\dot x_1^\mu \dot x_1^\nu  + \dot x_2^\mu \dot
x_2^\nu\right)+2 f_{\mu\nu} \dot x_1^\mu \dot x_2^\nu, \0\\
B &=& f_{\mu\nu} \left(\dot x_1^\mu \dot x_1^\nu  + \dot x_2^\mu \dot
x_2^\nu\right)+2 g_{\mu\nu} \dot x_1^\mu \dot x_2^\nu, \0
\ee
so that we have
\be
\widehat S[{\widehat x}] &=& \int  d\hat\tau \sqrt{\widehat g_{\mu\nu} \dot
{\widehat x}^\mu\dot {\widehat x}^\nu}\0\\
&=& 
{\frac 12 \left[
\int d\tau_1 \left( \sqrt{A+B} + \sqrt{A-B} \right)+\int d\tau_2
\left(\sqrt{A+B} -\sqrt{A-B}\right)\right]
}
\0\\&&{
+\frac {\gamma_5}2  \left[
\int d\tau_1 \left( \sqrt{A+B} - \sqrt{A-B} \right)+\int d\tau_2
\left(\sqrt{A+B} +\sqrt{A-B}\right)\right]
}
\label{gAB1}
\ee
Varying this action with respect to $\delta x^\lambda$ we
obtain the same eom (\ref{eomgeo}). This is due to \eqref{dxdx1int} and to the
fact that, the action is an analytic function of $\widehat x$, so that the
variation with respect to $\delta \widehat x^\lambda$ is the same as the
variation of  $\delta x_1^\lambda$.

***

Eventually we will set $x_2=0$ everywhere, but it is very convenient to keep the
axial-analytic notation as far as possible.

\subsection{Geodetic interval and distance}

The quantity
\be
\widehat E = E_1+ \gamma_5 E_2= \frac 12 \widehat g_{\mu\nu} \dot {\widehat
x}^\mu\dot {\widehat
x}^\nu\label{hatE}
\ee
is conserved as a function of $\hat t$. Since  $\widehat g_{\mu\nu} \dot
{\widehat
x}^\mu\dot {\widehat x}^\nu $ is constant for geodesics, we can write for the
arc length parameter $\widehat s$ 
\be
\frac {d\widehat s}{d\hat t} = \sqrt{\widehat g_{\mu\nu} \dot {\widehat
x}^\mu\dot
{\widehat
x}^\nu},\label{dsdt}
\ee
and 
\be
\widehat s-\widehat s' = \int_{\hat t'}^{\hat t} d\hat \tau \,\sqrt {2 \widehat
E}=\sqrt {2 \widehat E} \,
(\hat t-\hat t').\label{s1}
\ee
${\widehat s - \widehat s'}$ is the axial arc length along the geodesic between
$\widehat x$ and
$\widehat x'$. The half square of it
is called the {\it world function} and it is denoted
\be
\widehat \sigma(\widehat x, \widehat x')= \frac 12 (\widehat s-\widehat s')^2
=\widehat E (\hat t-\hat t')^2 =
(\hat t-\hat t')\int_{\hat t'}^{\hat t} \widehat E d\hat \tau\label{s2}
\ee
The main properties are
\be
\widehat \sigma_{;\mu} = \widehat \partial_\mu \widehat \sigma = (\hat t-\hat
t')\widehat
g_{\mu\nu} \dot{\widehat
x}^\nu\equiv -\widehat g_{\mu\nu} \widehat y^\nu\label{s3}
\ee
$\widehat y^\mu$ are the {\it normal coordinates} based at $\widehat x$. Using
(\ref{s2},\ref{s3}) one can see that
\be
\frac 12\widehat \sigma_{;\mu} \widehat \sigma_{;}{}^\mu = \widehat
\sigma\label{smus}
\ee
The subscript $_{;\mu}$ means the covariant derivative with respect to $\widehat
x^\mu$, while $_{;\mu'}$ means the covariant derivative with respect to
${\widehat x'}{}^{\mu'}$.

{\bf Remark 1}.   $\widehat \sigma = \sigma_1+\gamma_5 \sigma_2$, but
notice
that, even when we set $x_2=0$, we cannot infer that $\sigma_2=0$. This descends
from
eq.\eqref{dsdt}. Looking at \eqref{gAB1}, we see that $B$ does not vanish even
when $x_2^\nu=0$. As a consequence the
axial-imaginary part of \eqref{gAB} does not vanish,
so the axial-imaginary part of eq.\eqref{dsdt} will not automatically vanish
either.

\subsection{Normal coordinates}

Normal coordinates can be defined based at $x$ or at $x'$:
\be
\widehat y^{\mu'}(\widehat x',\widehat x) = (\hat t-\hat t') \frac {d \widehat
x^{\mu'}}{d\hat t'}\label{ncx'}
\ee
and
\be
\widehat y^{\mu}(\widehat x,\widehat x') = (\hat t'-\hat t) \frac {d \widehat
x^{\mu}}{d\hat t}\label{ncx}
\ee
The tangent vector $\frac {d \widehat x^{\mu}}{d\hat t}$ to the geodesic at
$\hat x$
satifies
\be
\frac {D}{d\hat t} \frac {d \widehat x^{\mu}}{d\hat t}=\frac {d^2\widehat
x^\mu}{d \hat t^2} +
{\widehat\Gamma}^\mu_{\nu\lambda} \frac {d \widehat x^{\nu}}{d\hat t}\frac {d
\widehat
x^{\lambda}}{d\hat t}=0\label{Dxdt}
\ee
and an analogous equation at $\hat x'$. Now we can write
\be
\widehat y^{\mu'}{}_{;\nu}(\hat x',\hat x)
\widehat y^\nu{(\hat x,\hat x')} &=&
(\hat t'-\hat t) \widehat
y^{\mu'}{}_{;\nu}(\widehat x',\widehat x) \frac {d\widehat x^\nu{(\hat t)}}{d
\hat
t}\0\\
&=& (\hat t'-\hat t) \frac d{d\hat t} \widehat y^{\mu'} (\widehat x',\widehat
x)=(\hat t'-\hat t)
\frac {d\widehat x^{\mu'}{(\hat t')}}{d\hat t'}
=-{\widehat y^{\mu'}}(\widehat x',\widehat x)\label{yy=y'}
\ee
Dividing by $\hat t-\hat t'$ the second and fourth terms and taking the
coincidence limit
$\widehat x'\to \widehat x$, one gets
\be
[\widehat y^{\mu'}{}_{;\nu}]\frac {d\widehat x^\nu}{d\hat t}= \frac {d\widehat
x^{\mu}}{d\hat t}
\quad\quad \rightarrow\quad\quad
[\widehat y^{\mu'}{}_{;\nu}] = \delta_\nu^\mu\label{ydelta1}
\ee
where $[X]$ denotes the result of the coincidence limit on the quantity $X$.  In
a similar way one
can prove
\be
[\widehat y^{\mu'}{}_{;\nu'}]\frac {d\widehat x^\nu}{d\hat t}&=&- \frac
{d\widehat
x^{\mu}}{d\hat t}
\quad\quad \rightarrow\quad\quad
[\widehat y^{\mu'}{}_{;\nu'}] = -\delta_\nu^\mu\label{ydelta2}
\ee
\be
[\widehat y^{\mu}{}_{;\nu}]\frac {d\widehat x^\nu}{d\hat t}&=&- \frac {d\widehat
x^{\mu}}{d\hat t}
\quad\quad \rightarrow\quad\quad
[\widehat y^{\mu}{}_{;\nu}] =- \delta_\nu^\mu\label{ydelta3}
\ee
\be
[\widehat y^{\mu}{}_{;\nu'}]\frac {d\widehat x^\nu}{dt}&=& \frac {d\widehat
x^{\mu}}{dt}
\quad\quad \rightarrow\quad\quad
[\widehat y^{\mu}{}_{;\nu'}] = \delta_\nu^\mu\label{ydelta4}
\ee

From \eqref{yy=y'} we get
\be
\widehat y^{\mu'}{}_{;\nu}\,\widehat y^\nu+ \widehat y^{\mu'}=0\label{y'y=y'}
\ee
In a similar way one derives also
\be
&&\widehat y^{\mu'}{}_{;\nu'}\,\widehat y^{\nu'}+ \widehat
y^{\mu'}=0\label{y'y'=y'}\\
&&\widehat y^{\mu}{}_{;\nu'}\,\widehat y^{\nu'}+ \widehat
y^{\mu}{}=0\label{yy'=y}\\
&&\widehat y^{\mu}{}_{;\nu}\,\widehat y^\nu+ \widehat y^{\mu}=0\label{yy=y}
\ee
For instance, differentiating \eqref{y'y'=y'} with respect to $\widehat
x^{\lambda'}$, one gets
\be
\widehat y^{\mu'}{}_{;\nu'\lambda'}\,\widehat y^{\nu'}+\widehat
y^{\mu'}{}_{;\nu'}\,\widehat
y^{\nu'}{}_{{;}\lambda'}+ \widehat y^{\mu'}{}_{;\lambda'}=0\0
\ee
taking the coincidence limit, and using \eqref{ydelta2}, one finds an identity,
because $[\widehat y^{\mu'}]=0$.
Differentiating another time with respect to $\widehat x^{\rho'}$ one gets
\be
[\widehat y^{\mu'}{}_{;\lambda'\rho'}]=0\label{ymlr}
\ee
Differentiating again with respect to $\widehat x^{\tau'}$ and using the Bianchi
identity for $\widehat
R^\mu{}_{\lambda\rho\tau}=R^{(1)\mu}{}_{\lambda\rho\tau}+\gamma_5
R^{(2)\mu}{}_{\lambda\rho\tau} $, one finds
\be
[\widehat y^{\mu'}{}_{;\lambda'\rho'\tau'}]=\frac 13 \left( \widehat
R^\mu{}_{\rho\lambda\tau}+ \widehat R^\mu{}_{\tau\lambda\rho}
\right)\label{ymlrt}
\ee
and, in a similar way,
\be
[\widehat y^{\mu'}{}_{;\lambda\rho\tau}]=\frac 13 \left( \widehat
R^\mu{}_{\lambda\rho\tau}+ \widehat R^\mu{}_{\rho\lambda\tau}
\right)\label{ymlrt1}
\ee
and
\be
[\widehat y^{\mu}{}_{;\lambda\rho\tau}]=\frac 13 \left( \widehat
R^\mu{}_{\tau\lambda\rho}+ \widehat R^\mu{}_{\rho\lambda\tau}
\right)\label{ymlrt2}
\ee

\subsection{Coincidence limits of $\widehat\sigma$}

Covariantly differentiating \eqref{smus} we get
\be
\widehat \sigma_{;\nu} = \widehat \sigma_{;\mu\nu} \widehat
\sigma_{;}{}^\mu\label{diffsigma1}
\ee
In the coincidence limit $[\widehat \sigma_{;\nu}]=0$. Therefore
\eqref{diffsigma1} is trivial in the coincidence limit.
Differentiating the first and last member of \eqref{s3} we get
\be
\widehat \sigma_{;\mu\lambda}= - \widehat g_{\mu\nu} \, \widehat
y^\nu{}_{;\lambda} \label{diffsigma2}
\ee
Using \eqref{ydelta3} one gets
\be
[\widehat \sigma_{;\mu\lambda}]=  \widehat g_{\mu\lambda} \label{diffsigma3}
\ee
Similarly
 \be
[\widehat \sigma_{;\mu\lambda'}]=-  \widehat g_{\mu\lambda} \label{diffsigma4}
\ee
Differentiating \eqref{diffsigma1} once more one gets
\be
\widehat \sigma_{;\nu\lambda} = \widehat \sigma_{;\mu\nu\lambda}\, \widehat
\sigma_{;}{}^\mu + \widehat \sigma_{;\mu\nu}\, \widehat \sigma_{;\lambda}^\mu\0
\ee
which, in the coincidence limit, using the previous results, yields an identity.
Differentiating it again
\be
\widehat \sigma_{;\nu\lambda\rho} = \widehat \sigma_{;\mu\nu\lambda\rho}\,
\widehat \sigma_{;}{}^\mu + \widehat \sigma_{;\mu\nu\lambda}\, \widehat
\sigma^\mu_{;\rho}
+ \widehat \sigma_{;\mu\nu\rho}\, \widehat \sigma_{;\lambda}^\mu + \widehat
\sigma_{;\mu\nu}\, {{\widehat \sigma}_{;}{}^\mu}_{\lambda\rho}\label{diffsigma5}
\ee
In the coincidence limit this becomes
\be
[\widehat \sigma_{;\nu\lambda\rho}]= [\widehat \sigma_{;\rho\nu\lambda}]+
[\widehat \sigma_{;\lambda\nu\rho}] +[\widehat
\sigma_{;\nu\lambda\rho}]\label{diffsigma6}
\ee
Since $\widehat \sigma$ is a biscalar we have
\be
[\widehat \sigma_{;\nu\lambda\rho}]= [\widehat \sigma_{;\nu\rho\lambda}]+
\widehat R_{\rho\lambda\nu}{}^\tau [\widehat \sigma_{;\tau}] =
[\widehat \sigma_{;\rho\nu\lambda}]\label{diffsigma7}
\ee
Therefore
\be
[\widehat \sigma_{;\rho\nu\lambda}]= [\widehat \sigma_{;\lambda\nu\rho}] =
[\widehat \sigma_{;\nu\lambda\rho}]=0\label{diffsigma8}
\ee

Differentiating \eqref{diffsigma5} once more and taking the coincidence limit
one gets
{
\be
[\widehat\sigma_{;\nu\lambda\rho\tau}] = -\frac 13 \left(\widehat
R_{\nu\tau\lambda\rho}+\widehat R_{\nu\rho\lambda\tau}\right)\equiv\widehat
S_{\nu\lambda\rho\tau}\label{diffsigma9}
\ee}
where $\widehat R_{\nu\tau\lambda\rho}= \widehat g_{\nu\mu} \widehat
R^\mu{}_{\tau\lambda\rho}$.
Differentiating once more
\be
[\widehat\sigma_{;\nu\lambda\rho\sigma\tau}]= \frac 34 \left(\widehat S_{
\nu\lambda\sigma\tau;\rho}+\widehat S_{ \nu\lambda\sigma\rho;\tau}
+\widehat S_{ \nu\lambda\tau\rho;\sigma}\right)\label{diffsigma10}
\ee

We will need also the coincidence limits of tensors covariantly differentiated
with respect to a primed index $\nu'$. In general
\be
[t_{\mu_1\ldots\mu_k;\nu'}] =[t_{\mu_1\ldots\mu_k}]_{;\nu} -
[t_{\mu_1\ldots\mu_k;\nu}]\label{tnu'}
\ee
So
\be
[\widehat \sigma_{;\mu\nu'}]&=&[ \widehat \sigma_{;\mu}]_{;\nu} - [\widehat
\sigma_{;\mu\nu}]=-\widehat g_{\mu\nu}\label{sigmanu'}
\ee
\be
[\widehat \sigma_{;\mu\nu'\lambda}]&=& [\widehat \sigma_{;\mu\lambda\nu'}]=
[\widehat \sigma_{;\mu\lambda}]_{;\nu}
- [\widehat \sigma_{;\mu\lambda\nu}]=0\label{sigmamn'l}
\ee
\be
[\widehat \sigma_{;\mu\nu'\lambda\rho}]&=& [\widehat
\sigma_{;\mu\lambda\rho\nu'}]= [\widehat \sigma_{;\mu\lambda\rho}]_{;\nu}-
[\widehat \sigma_{;\mu\lambda\rho\nu}]=-[\widehat \sigma_{;\mu\lambda\rho\nu}]
=- \widehat S_{\mu\lambda\rho\nu}\label{sigmamlrn}
\ee
and
\be
[\widehat \sigma_{;\mu\nu'\lambda\rho\sigma}]&=& [\widehat
\sigma_{;\mu\lambda\rho\sigma\nu'}]=
[\widehat \sigma_{;\mu\lambda\rho\sigma }]_{;\nu}-[\widehat
\sigma_{;\mu\lambda\rho\sigma\nu}]= \frac 14 \widehat
S_{\mu\lambda\rho\sigma;\nu}
-\frac 34 \left(\widehat S_{\mu\lambda\nu\rho;\sigma}+ \widehat
S_{\mu\lambda\sigma\nu;\rho}\right)\label{sigmamlrsn}
\ee
{Similarly, one obtains}
\be
{[\widehat \sigma_{;\mu}{}^\mu{}_\nu{}^\nu{}_\rho{}^\rho]= 
-\frac 85 R_{;\mu}{}^\mu
+\frac 4{15} \widehat R_{\mu\nu}\widehat R^{\mu\nu} -\frac 4{15} \widehat
R_{\mu\nu\lambda\rho}\widehat R^{\mu\nu\lambda\rho}}\0\label{sigmammnnrr}
\ee
\be
{[\widehat \sigma_{;\mu}{}_\nu{}^\nu{}_\rho{}^\rho{}^\mu]= 
-[\widehat \sigma_{;\mu}{}^{\mu'}{}_\nu{}^\nu{}_\rho{}^\rho]= 
\frac 25 R_{;\mu}{}^\mu
-\frac 1{15} \widehat R_{\mu\nu}\widehat R^{\mu\nu} -\frac 4{15} \widehat
R_{\mu\nu\lambda\rho}\widehat R^{\mu\nu\lambda\rho}}\0\label{sigmamnnrrm}
\ee

\subsection{Van Vleck-Morette determinant }

The Van Vleck-Morette determinant in MAT is defined by
\be
\widehat D(\widehat x,\widehat x') = \det
(-\widehat\sigma_{{;}\mu\nu'})\label{VVM}
\ee
$\widehat D(\widehat x,\widehat x')$ is a bidensity of weight 1 both at
$\widehat x$ and $\widehat x'$.
Later on we will need a bidensity of weight 0:
\be
\widehat \Delta (\widehat x,\widehat x')= \frac 1{\sqrt{\widehat g(\widehat
x)}}
\widehat D(\widehat x,\widehat x') \frac 1{\sqrt{\widehat g(\widehat
x')}}\label{Delta}
\ee
The VVM determinant also satisfies (for 4 dimensions)
\be
(\widehat D(\widehat x,\widehat x') \widehat\sigma^{{;}\mu})_{;\mu} = 4 \widehat
D(\widehat x,\widehat x')
\label{DeltaTrace}
\ee
In the coincidence limit
\be
[\widehat \Delta^{\frac 12}_{;\lambda}] = [\widehat g^{-\frac 14}(\widehat x)
\sqrt{{\widehat D(\widehat x,\widehat x')}} \frac 12
\left( \widehat \sigma^{-1}{}^{{\mu\nu'}}\widehat
\sigma_{;\mu\nu'\lambda}\right)
\widehat g^{-\frac 14}(\widehat x') ]
= \frac{1}{2}[\widehat \sigma^\mu_{;\mu\lambda}]=0\label{VVM2}
\ee
We need to compute the covariant derivatives of ${\widehat
\sigma}^{-1}{}^{\mu\nu'}\equiv \{\widehat \sigma^{-1}_{;\mu\nu'}\}$. The latter
is defined as
\be
\widehat \sigma^{-1}{}^{\mu\nu'} \widehat \sigma_{;\nu'\lambda} =
\delta^\mu_\lambda \label{inversesigma}
\ee
Differentiating this relation once, twice {and thrice} one gets
\be
&&[\widehat \sigma^{-1}{}^{\mu\nu'}{}_{;\lambda}]=0,\label{inversesigma1}\0
\\&&
[\widehat
\sigma^{-1}{}_{\mu\lambda'}{}_{;\rho\sigma}]=-[\widehat\sigma_{
;\mu'\lambda\rho\sigma}]=[\widehat\sigma_{;\lambda\rho\sigma\mu}] =
\widehat S_{\lambda\rho\sigma\mu}\label{inversesigma2}
\ee
and
\be
{[\widehat
\sigma^{-1}{}_{\mu\lambda'}{}_{;\rho\sigma\tau}]
=-[\widehat\sigma_{;\lambda\mu'\rho\sigma\tau}]\label{inversesigma3}
=\frac 14 \widehat
S_{\mu\rho\sigma\tau;\lambda}
-\frac 34 \left(\widehat S_{\mu\rho\lambda\sigma;\tau}+ \widehat
S_{\mu\rho\tau\lambda;\sigma}\right)}
\ee

Differentiating once more one gets
\be
[\widehat \Delta^{\frac 12}_{;\lambda\rho}]= \frac 16 \widehat g^{\mu\nu}\left(
\widehat R_{\mu\nu\lambda\rho}
+ \widehat R_{\mu\lambda\nu\rho}\right)=
\frac 16 \widehat g^{\mu\nu}\widehat g_{\mu\sigma}  \widehat
R^\sigma{}_{\lambda\nu\rho}
=\frac 16 \left(R^{(1)}_{\lambda\rho} + \gamma_5
R^{(2)}_{\lambda\rho}\right)\label{VVM3}
\ee
and
\be
[\widehat \Delta^{\frac 12}_{;\lambda\rho\sigma}]= \frac 1{12} \left(\widehat
R_{ \lambda\rho;\sigma} + \widehat R_{\rho\sigma;\lambda}+
\widehat R_{ \sigma\lambda;\rho}\right)\label{VVM4}
\ee
Finally
\be
[\widehat \Delta^{\frac 12}_{;\mu}{}^\mu{}_\nu{}^\nu]=  {+}\frac 1{5} \widehat
R_{;\mu}{}^\mu+\frac 1{36} \widehat R^2
-\frac 1{30} \widehat R_{\mu\nu}\widehat R^{\mu\nu} + \frac 1{30} \widehat
R_{\mu\nu\lambda\rho}\widehat R^{\mu\nu\lambda\rho}\label{VVM5}
\ee

\subsection{The geodetic parallel displacement matrix}

The geodetic parallel displacement matrix $\widehat G^\mu{}_{\nu'}(\widehat
x,\widehat x')$ is needed in order
to parallel displace vectors  from one end to the other of the geodetic
interval. It is defined by
\be
[\widehat G^\mu{}_{\nu'}]=\delta^\mu_\nu , \quad\quad \widehat
G^\mu{}_{\nu';\lambda}\widehat \sigma^{;\lambda}=0\label{GPD}
\ee
The second condition means that the covariant derivative of $\widehat
G^\mu{}_{\nu'}$ vanishes in directions
parallel to the geodesic. Since tangents to the geodesics are self-parallel, it
follows that
\be
&&\widehat G_\mu{}^{\nu'} \,\widehat \sigma_{;\nu'}= - \sigma_{;\mu} ,\quad\quad
\widehat \sigma_{;\mu} \,\widehat G^\mu{}_{\nu'}=-\widehat
\sigma_{;\nu'}\label{GPD1}\\
&& \widehat G_{\mu\nu'}= \widehat G_{\nu'\mu}, \quad\quad \widehat
\sigma_{;}{}^{\lambda'}\widehat G^{\mu} {}_{\nu';\lambda'}=0\0\\
&& \widehat  G_{\mu}{}^{\nu'}\widehat G_{\nu'}{}^{\lambda}=
\delta_{\mu}^\lambda\0
\ee

The analogous parallel displacement for spinors is denoted $I(x,x')$: the object
$I(x,x')\psi(x')$ is the
spinor $\psi(x)$ obtained by parallel displacement of $\psi(x')$ along the
geodesic from $x'$ to $x$.
 It is a bispinor quantity satisfying
\be
\widehat\sigma_{;}{}^\mu \widehat I_{;\mu}=0 ,\quad\quad [\widehat I]={\bf
1}\label{I}
\ee
and ${\bf 1}$ is the identity matrix in the spinor space.
Differentiating \eqref{I} once we get $[\widehat I_{;\mu}]=0$. Differentiating
twice we get
{
\be
[\widehat I_{;(\mu\nu)}]=0,\label{I2}
\ee}
while
\be
\widehat I (x,x')_{;\mu\nu}-\widehat I(x,x')_{;\nu\mu} = -\frac 12 \left( d
\widehat \Omega+\widehat \Omega \widehat \Omega\right)_{\mu\nu}
\widehat I(x,x')= -\frac 12 \widehat{\cal R}_{\mu\nu} I(x,x')\label{I3}
\ee
where $\widehat {\cal R}_{\mu\nu}= {\widehat R}_{\mu\nu}{}^{ab}
\Sigma_{ab}$.
So
{
\be
[\widehat I (x,x')_{;[\mu,\nu]}] = [\widehat I (x,x')_{;\mu\nu}] = -\frac 14
\widehat{\cal R}_{\mu\nu}\label{I4}
\ee}

Proceeding with the differentiations of \eqref{I} we find
{
\be
[\widehat I_{;\nu\lambda\rho}] + [\widehat I_{; \lambda\nu\rho}] + [\widehat
I_{;\rho\lambda\nu}] =0\label{I5}
\ee}
Now
\be
[\widehat I_{;\nu\lambda\rho}] - [\widehat I_{;\nu\rho\lambda}]= \frac 12
\widehat{\cal R}_{\rho\lambda} [\widehat I_{;\nu}]=0\label{I6}
\ee
and
\be
3[\widehat I_{;\nu\lambda\rho}] = \frac 12 \widehat{\nabla}_\rho \widehat {\cal
R}_{\lambda\nu} +\frac 12   \widehat{\nabla}_\lambda \widehat{\cal R}_{\rho\nu}
\label{I7}
\ee
In particular
\be
[\widehat I_{;\nu}{}^\nu{}_\rho ]= \frac 16 \widehat{\nabla}^\nu   \widehat
{\cal
R}_{\rho\nu}\label{I8}
\ee
{
Differentiating \eqref{I} once more with respect to $x^\sigma$, using
\eqref{diffsigma9} and then
contracting with $\widehat g^{\nu\lambda} \widehat g^{\sigma \rho}$ we find,
after simplifying,
\be
[\widehat I_{;\mu}{}^{\mu}{}_{\nu}{}^{\nu}]+ [\widehat I_{;\mu\nu}{}^{\nu\mu}]=0
\label{I9}
\ee
A contraction with $\widehat g^{\nu\sigma} \widehat g^{\lambda \rho}$ gives:
\be
[\widehat I_{;\mu\nu}{}^{\nu\mu}] + 2 [\widehat I_{;\mu\nu}{}^{\mu\nu}] +
[\widehat I_{;\mu}{}^{\mu}{}_{\nu}{}^{\nu}] =0
\label{I10}
\ee
Using \eqref{I3}, we get
\be
[\widehat I_{;\sigma \rho\mu\nu}] = [\widehat{\nabla}_\nu\widehat{\nabla}_\mu
(\widehat{I}_{;\sigma\rho})] = -\frac{1}{2}\widehat{\cal R}_{\sigma\rho;\mu\nu}
+ \frac{1}{8}\widehat{\cal R}_{\sigma\rho}\widehat{\cal R}_{\mu\nu} + [\widehat
I_{;\rho\sigma \mu\nu}]
\ee
Contracting with $\widehat g^{\mu\sigma} \widehat g^{\nu \rho}$ gives
\be
[\widehat I_{;\mu\nu}{}^{\mu\nu}] = 0 + \frac{1}{8}\widehat{\cal
R}_{\mu\nu}\widehat{\cal R}^{\mu\nu} + [\widehat I_{;\mu\nu}{}^{\nu\mu}]
\ee
since by Walker's identity
\be
\widehat\nabla_\rho \widehat\nabla_\lambda \widehat{\cal R}^{\rho\lambda}=
0 \label{nablanablaR}
\ee
Finally, by using \eqref{I9}, \eqref{I10}, one gets
\be
[\widehat I_{;\nu}{}^\nu{}_\rho{}^\rho]=  \frac 18 \widehat{\cal
R}_{\rho\lambda} \widehat{\cal R}^{\rho\lambda}\label{I11}
\ee}

\section{Fermions in MAT background}

The action of a fermion interacting with a metric and an axial tensor is
\be
\widehat S&=&\int d^4\widehat x \, \left(i\overline {\psi}
\sqrt{\overline{\widehat
g}}\gamma^a\widehat
e_a^\mu
\left(\partial_\mu+\frac 12 \widehat\Omega_\mu \right)\psi\right)(\widehat x)\label{axialaction}
\\
&{=}&  \int d^4\widehat x \,\left(  i\overline {\psi} \sqrt{\overline{\widehat
g}
}\gamma^a(\tilde e_a^\mu+\gamma_5
\tilde c_a^\mu)  \left(\partial_\mu +\frac 12 \left(\Omega^{(1)}_\mu+\gamma_5
\Omega^{(2)}_\mu\right) \right)\psi\right)(\widehat x)   \0\\
&=&\int d^4\widehat x \,\left(i \overline {\psi} \sqrt{\overline{\widehat g}
}(\tilde
e_a^\mu-\gamma_5 \tilde
c_a^\mu)\left[\frac  12 \gamma^a {\stackrel{\leftrightarrow}
{\partial}}_\mu + {\frac 14} \left( \gamma^a \widehat\Omega_\mu +
\overline{\widehat
 \Omega}_\mu
\gamma^a\right) \right]\psi\right)(\widehat x)  \0\\
&=& \int d^4\widehat x \, \left(i\overline {\psi} \sqrt{\overline {\widehat
g}}(\tilde e_a^\mu-\gamma_5 \tilde
c_a^\mu)\left[\frac  12 \gamma^a {\stackrel{\leftrightarrow}
{\partial}}_\mu \psi+ \frac i4 \gamma_d \epsilon^{dabc}\widehat \Omega_{\mu bc}
\gamma_5\right]\psi \right)(\widehat x)  \0
\ee
It must be noticed that this action takes axial-real values
\footnote{ One could consider also an axial complex action, 
but for our purposes this is a useless complication. That is why we use the
notation $\psi$ 
instead of $\widehat \psi $.}. The field $\psi(\widehat x)$
can be understood, classically, as a series of powers of $\widehat x$ 
applied to constant spinors on their right and the symmetry transformations act
on it 
from the left. The analogous definitions for $\psi^\dagger$ are obtained via
hermitean conjugation.  
In the second line it is stressed that the action contains also an axial part.
It is understood that $\partial_\mu=\frac {\partial}{\partial \widehat x^\mu}$
applies only to $\psi$ or $\overline \psi$, as indicated, and
$\overline{\widehat g}$ denotes, as usual, the
axial-complex conjugate of $\widehat g$.

A few comments are in order.
As was explained in \cite{MAT1}, the density  $\sqrt{\overline{\widehat g}}$
must be inserted
between $\overline \psi$ and $\psi$, due to the presence in it of the $\gamma_5$
matrix.
Moreover one has to take into account that the kinetic
operator contains a $\gamma$ matrix that anticommutes with $\gamma_5$. Thus, for
instance, using ${\widehat D}_\lambda \widehat g_{\mu\nu}=0$ and $({\widehat
D}_\lambda +\frac 12
\widehat \Omega_\lambda) \widehat e=0$, where ${\widehat D}=\partial+\widehat
\Gamma$, one gets
 \be
{\overline {\psi}} \gamma^a\widehat e_a^\mu
\left(\partial_\mu+\frac 12 \Omega_\mu \right)\psi = \overline {\psi} (\overline
{\widehat D}_\mu
+\frac 12  \overline {\widehat\Omega}_\mu) \gamma^a\widehat
e_a^\mu\psi\label{equation}
\ee
We recall again that a bar denotes axial-complex conjugation, i.e. a sign
reversal in front of
each $\gamma_5$ contained in the expression, for instance   $ \overline{\widehat
 \Omega}_\mu=
\Omega_\mu^{(1)}-\gamma_5 \Omega_\mu^{(2)}$.

To obtain the two last lines in \eqref{axialaction}
one must use \eqref{Gammamumunu} and \eqref{equation}.

\subsection{{Classical Ward identities}}

Let us consider AE (axially extended) diffeomorphisms first, \eqref{axialdiff}.
It is not hard to prove that the action \eqref{axialaction} is invariant under
these transformations.
Now, define the full MAT e.m. tensor by means of
\be
{\bf T}^{\mu\nu} = \frac 2{\sqrt{\widehat g}} \frac {\stackrel{\leftarrow}
{\delta}{\widehat S}}
{\delta \widehat g_{\mu\nu}} \label{fullem0}
\ee
This formula needs a comment, since $\sqrt{\widehat g}$ contains $\gamma_5$. To
give a
meaning to it we understand that
the operator $\frac 2{\sqrt{\widehat g}} \frac {\stackrel{\leftarrow} {\delta} }
{\delta \widehat g_{\mu\nu}}$  in the RHS acts on the operatorial expression,
say
${\cal O}{\sqrt{\widehat g}}$, which is inside the scalar product 
$\overline \psi
{\cal O}\sqrt{\widehat g} \psi$.
Moreover the functional derivative acts from the right of the action.
Now the conservation law under diffemorphisms is
\be
0=\delta_{\widehat \xi} S &=& \int \overline \psi \frac { \stackrel{\leftarrow}
{\delta}
{\cal O}} {\delta \widehat g_{\mu\nu}}\delta \widehat g_{\mu\nu}\psi
=\int \overline \psi\frac {\stackrel{\leftarrow} {\delta} {\cal O}} {\delta
\widehat g_{\mu\nu}} \left({\widehat D}_\mu \widehat\xi_\nu+ {\widehat D}_\nu
\widehat\xi_\mu\right)\psi
\0\\
&=& -2 \int \overline \psi \frac {\stackrel{\leftarrow} {\delta} {\cal O}}
{\delta \widehat g_{\mu\nu}}  {\stackrel{\leftarrow} {\widehat D}}_\mu
\widehat\xi_\nu\psi\label{deltaXi}
\ee
where ${\widehat D}$ acts (from the right) on everything except the parameter
$\widehat\xi_\nu$. Differentiating with respect to the arbitrary
parameters $\xi^\mu$ and $\zeta^\nu$ we obtain two conservation laws involving
the two tensors
\be
T^{\mu\nu}&=&{2}  \overline \psi \frac {\stackrel{\leftarrow} {\delta} {\cal O}}
{\delta \widehat g_{\mu\nu}}\psi\label{Tmunu}\\
T_5^{\mu\nu} &=& {2} \overline \psi \frac {\stackrel{\leftarrow} {\delta} {\cal
O}}
{\delta \widehat g_{\mu\nu}}\gamma_5\psi\label{T5munu}
\ee
To give a less abstract idea of these tensors, at the lowest order (flat
background) and setting $x_2^\mu=0$, they are given by
\be
T^{\mu\nu}\approx T_{flat}^{\mu\nu}=-\frac i4 \left(\overline {\psi } \gamma^\mu
{\stackrel{\leftrightarrow}{\partial^\nu}}\psi
 + \mu\leftrightarrow \nu \right),\label{Tmunu0}
\ee
and
\be
T_5^{\mu\nu}\approx T_{5 flat}^{\mu\nu}=\frac i4 \left(\overline {\psi}\gamma_5
\gamma^\mu  {\stackrel{\leftrightarrow}{\partial^\nu}}\psi
 + \mu\leftrightarrow \nu \right),\label{T5munu0}
\ee

Repeating the same derivation for the axial complex Weyl transformation one can
prove that, assuming for the fermion field the transformation rule
\be
\psi \rightarrow e^{-\frac 32 (\omega+\gamma_5 \eta)} \psi,\label{Weylpsi}
\ee
\eqref{axialaction} is invariant, and obtain the Ward identity
\be
0= \int \overline \psi \frac {\stackrel{\leftarrow} {\delta} {\cal O}} {\delta
\widehat g_{\mu\nu}} \widehat g_{\mu\nu} \,(\omega+\gamma_5 \eta)\psi
\label{WeylWI}
\ee
One gets in this way two WI's
\be
{\ET}(x)&\equiv&T^{\mu\nu} g_{\mu\nu} + T_{5}^{\mu\nu} f_{\mu\nu}=0,
\label{WardWeyl1}\\
{\ET}_5(x)&\equiv &T^{\mu\nu} f_{\mu\nu} + T_{5}^{\mu\nu} g_{\mu\nu}=0,
\label{WardWeyl2}
\ee

\subsection{A more precise formula for the e.m. tensor}

In our calculation a more explicit formula of the e.m. tensor is needed. 
The e.m. tensor is defined by
\be
{\bf T}^{\mu\nu}  = \frac 2{\sqrt{{\widehat g}}} \frac {\stackrel{\leftarrow}
{\delta} {\widehat S}}
{\delta \widehat g_{\mu\nu}} = \frac 12 \left( {\bf T}_a^\mu \widehat e^{a\nu} +
 {\bf
T}_a^\nu \widehat e^{a\mu}\right)\label{fullem}
\ee
where
\be
{\bf T}_a^\mu = \frac 1{\sqrt{|{\widehat g}|}} \frac {\stackrel{\leftarrow}
{\delta}{\widehat S}}{\delta \widehat  e^a_\mu}\label{halfem}
\ee
Let us prove first that the functional derivative of $\widehat \Omega_m$ does
not contribute to the e.m. tensor.
Consider the general variational formula
\be
\delta \widehat\Omega_\mu^{bc} &=& \frac 12 \widehat e^{b\nu}
\left(\widehat\nabla_\mu (\delta \widehat e_\nu^c)
- \widehat\nabla_\nu (\delta \widehat e_\mu^c)\right)-
\frac 12 \widehat e^{c\nu} \left(\widehat\nabla_\mu (\delta \widehat e_\nu^b)
 - \widehat\nabla_\nu (\delta \widehat e_\mu^b)\right)\0\\
&&+ \frac 12 \widehat e^{b\nu} \widehat e^{c\lambda}
\left(\widehat\nabla_\lambda (\delta \widehat e_\nu^e)
 - \widehat\nabla_\nu (\delta \widehat e_\lambda^e)\right)
\widehat e_{e\mu}\label{deltaOmega}
\ee
where $\widehat\nabla$ denotes the covariant derivative such that
 $\widehat\nabla_\mu \widehat e^a_\lambda=0$. After some algebra one gets
\be
\gamma_d\, \epsilon^{dabc}\,\widehat e^\mu_a \,\delta\widehat \Omega_{\mu bc}
= \gamma_d \,\epsilon^{dabc}\,\widehat e^\mu_a \widehat e_b^\nu \nabla_\mu
\delta
e_{c\nu}
\label{gammadepsilon}
\ee
Now use this and
\be
\frac {\delta \widehat e^a_\mu(x)}{\delta \widehat e^b_\nu (y)} = \delta^a_b
\delta^\nu_\mu \delta(x,y)\0
\ee
and insert them into the definition \eqref{fullem}. The relevant contribution is
\be
{\bf T}_\Omega^{\lambda\rho}  &=&\frac 12 \left( {\bf T}_a^\lambda \widehat
e^{a\rho}
+  {\bf T}_a^\rho \widehat e^{a\lambda}\right)_\Omega\label{TOmega}\\
&\equiv& \frac 18 \int \overline \psi  \gamma_d \epsilon^{dabc} \widehat e^\mu_a
\left(\frac {\delta\widehat \Omega_{\mu bc}} {\delta \widehat e^e_\lambda}
\widehat e^{e\rho}+
\frac {\delta\widehat \Omega_{\mu bc}} {\delta \widehat e^e_\rho} \widehat
e^{e\lambda}\right) \gamma_5 \psi\0\\
&=& \frac 18 \int \overline \psi  \gamma_d \epsilon^{dabc} \widehat
e^\mu_a\left( \widehat e_b^\lambda \,\widehat e_c^\rho \widehat \nabla_\mu
\delta(x,y)
+ \widehat e_b^\rho \,\widehat e_c^\lambda \widehat\nabla_\mu \delta(x,y)\right)
\gamma_5 \psi=0\0
\ee
Therefore the only contribution to the em tensor comes from the variation of the
first $\widehat e^m_a $ factor in \eqref{axialaction}.
The result is
\be
{\bf T}^{\lambda\rho} & =& -\frac i2 \overline\psi\widehat \gamma^\lambda
\widehat g^{\rho\mu} \left(\partial_\mu
+\frac 12 \widehat \Omega_\mu\right) + (\lambda\leftrightarrow \rho)
=- \frac i2 \overline\psi\widehat \gamma^\lambda \widehat\nabla^\rho
\psi+(\lambda\leftrightarrow \rho) \label{Tlambdarho}
\ee
where $\widehat\gamma^\lambda= \gamma^a \widehat e^\lambda_a$.

It is useful to write it as a trace
\be
{\bf T}^{\lambda\rho}(x)= \frac i2\tr\left(\eta \widehat \gamma^{(\lambda}
\widehat\nabla^{\rho)} \psi(x) \psi^\dagger(x)\right)
= \frac i4 \tr \left(\eta \widehat \gamma^{(\lambda} [\widehat\nabla^{\rho)}
\psi(x), \psi^\dagger(x)]\right)\label{Tsplit}
\ee
where $\eta\equiv \gamma_0$, the flat gamma matrix. The commutator is
interpreted as
\be
[\widehat\nabla^{\rho} \psi, \psi^\dagger](x)=\frac 12 \lim_{x'\to x} \left([
\widehat\nabla^{\rho} \psi(x), \psi^\dagger(x')]
+[ \widehat\nabla^{\rho} \psi(x'), \psi^\dagger(x)] \right)\label{pointsplit}
\ee
Inserting \eqref{Tsplit} in the path integral it becomes
\be
\langle\!\langle {\bf T}^{\lambda\rho}(x)\rangle\!\rangle = \frac i8 \lim_{x'\to
x}  \, \tr \left(\eta  \widehat \gamma^{(\lambda}\left(
\widehat\ES^{(1);\rho)} (x,x') - \widehat\ES^{(1);\rho')}
(x,x')\right)\right)\label{TS1}
\ee
where $\widehat\ES^{(1)}$ is the Hadamard function
\be
\widehat\ES^{(1)} (x,x')= \langle\!\langle [\psi(x) , \psi^\dagger (x')
]\rangle\!\rangle \label{Hadamard}
\ee
This leads to Christensen's method, \cite{Christ1,Christ2}, to compute the
energy-momentum tensor and related quantities, such as trace anomalies. We will
not pursue this point of view here although it could be done. It is in fact
strictly connected with the main approach we will follow later on, which we
consider simpler. They are both based on fermion propagators such as 
$\widehat\ES^{(1)} (x,x')$. A discussion of fermion propagators and their
properties in a MAT background is presented in Appendix \ref{sec:prop}.

\subsection{The Dirac operator and its inverse}

In the action \eqref{axialaction} the Dirac operator is
\be
\widehat F=i \widehat \gamma \! \cdot \! \widehat \nabla= i \widehat
\gamma^\mu \widehat\nabla_\mu = i  \gamma^a \widehat e^\mu_a
\widehat\nabla_\mu\equiv \gamma^a \, \widehat F_a\label{hatF}
\ee
where the $\widehat \nabla$ operator is, schematically, $\widehat D+ \frac 12
\widehat\Omega$ and satisfies $\widehat\nabla_\mu \widehat e^a_\nu=0$. 

Under AE diffeomorphisms $\psi$ transforms as:  $\delta_{\hat\xi} \psi =
\widehat \xi \!\cdot\! \partial\psi$, while
\be
\delta_{\hat\xi} \left(i\widehat \gamma\!\cdot\!\widehat \nabla \psi\right) =
\overline{\widehat \xi} \!\cdot\! \partial \left(i\widehat
\gamma\!\cdot\!\widehat \nabla \psi\right) \label{deltahatxiDirac}
\ee
Under AE Weyl transformation $\widehat F$ transform as
\be
\delta_{\hat \omega} \widehat F = -\frac 12 \gamma^a \{ \widehat F_a, \widehat
\omega\}\label{hatFWeyl}
\ee
and it has the following hermiticity property
\be
{\widehat F}^\dagger= \eta \widehat F \eta\label{Fdagger}
\ee
where $\eta =\gamma_0$ and $\gamma_0$ is the nondynamical (flat) gamma matrix.
To obtain \eqref{Fdagger} use $\widehat \Omega ^\dagger = - \eta \overline{
\widehat \Omega}^\dagger \eta $, etc.

Integrating out the fermion field in \eqref{axialaction} means, roughly
speaking, evaluating the determinant of the Dirac operator $\widehat F$. This is
however not what we need. First, because the log of the determinant is formally
the trace of the log of $\widehat F$; taking this trace means integrating
over spacetime and tracing over the gamma matrices: this would suppress any
explicit $\gamma_5$ dependence and, thus, any axial splitting. Second, because
$\widehat F$ is local, while, in order to exploit a coincidence limit (in order
to guarantee covariance), we need a bilocal quantity. This quantity exists, it
is the inverse of $\widehat F$: the fermion propagator. The Schwinger-DeWitt
method is based on it. Let us explain this approach, adapting it to MAT.

One starts from
\be
\widehat G(\widehat x,\widehat x') = \langle 0| {\cal T} \psi(\widehat x)
\psi^\dagger (\widehat x')|0\rangle \label{hatFeyn}
\ee
which satisfies
\be
i \sqrt{\widehat g} \eta\,\widehat \gamma^\mu \widehat\nabla_\mu
\widehat G(\widehat x,\widehat x') = -{\bf 1}{\mathbf{\delta }}(\widehat
x,\widehat x')\label{hatFpropa}
\ee
where ${\bf 1}$ is the unit matrix in the spinor space.
$\widehat G$ is not yet what we need. The Schwinger-DeWitt method requires a
quadratic operator and, in addition, we must get rid of the $\gamma$ matrices,
except $\gamma_5$. This is achieved
with the ansatz
\be
\widehat G(x,x') = -i \overline{ \widehat  \gamma}^\mu\overline{
\widehat\nabla}_\mu \overline{\widehat \EG}(x,x') \eta^{-1}
\label{hatansatz1}
\ee
\vskip 1cm
{\bf Remark 2.} {\footnotesize  {Why the ansatz \eqref{hatansatz1}}

In ordinary gravity, from the diff invariance of the fermion action, we can
extract the transformation rule
\be
\delta_\xi   \left(i\gamma^\mu \nabla_\mu\psi\right) =\xi\!\cdot\! \partial
\left(i  \gamma\!\cdot\! \nabla\psi\right)   \label{deltaxiDirac}
\ee
while $\delta_\xi \psi =\xi\!\cdot\! \partial\psi$.  Therefore it makes sense to
apply $ \gamma\!\cdot\! \nabla$ to $ \gamma\!\cdot\! \nabla\psi$,
because the latter transforms as $\psi$. This allows us to define the square of
the Dirac operator:
\be
F^2\psi = \left( i \gamma\!\cdot\! \nabla\right)^2 \psi\label{Gpsi}
\ee
It is not possible to repeat the same thing for MAT because of
\eqref{deltahatxiDirac}, from which we see that $\left(i\widehat
\gamma\!\cdot\!\widehat \nabla \psi\right) $ does not
transform like $\psi$, and an expression like $
\left(i\widehat \gamma\!\cdot\!\widehat \nabla\right)^2 \psi$ would break
general covariance. Noting that
\be
\delta_{\hat\xi} \left(i\overline{\widehat \gamma}\!\cdot\!\overline{\widehat
\nabla} \psi\right) =
{\widehat \xi} \!\cdot\! \partial \left(i\overline{\widehat
\gamma}\!\cdot\!\overline{\widehat \nabla} \psi \right) \label{deltahatxiDirac2}
\ee
when $\delta_{\overline{\widehat \xi}}\psi= \overline{\widehat \xi} \!\cdot\!
\partial\psi$,
we will consider instead the {\it covariant} quadratic operator
\be
\left(i\overline{\widehat \gamma}\!\cdot\!\overline{\widehat
\nabla}\right)\,\left(i {\widehat \gamma}\!\cdot\! {\widehat
\nabla}\right)\psi\label{covDiracsquare}
\ee}
\vskip 1cm

Let us quote next a few useful identities.
\be
\overline{\widehat \nabla}_\mu {\widehat \gamma}_\nu -{\widehat
\gamma}_\nu{\widehat \nabla}_\mu =
\gamma^a \left( \partial_\mu\, \widehat e_{a\nu} - \widehat
\Gamma_{\mu\nu}^\lambda {\widehat e_{a\lambda}}  +\frac 12 \widehat \Omega_{\mu
ab}\,\widehat e ^b_\nu\right)=0\label{metricity}
\ee
because of metricity, and
\be
\overline{\widehat \nabla}_\mu {  \gamma}^a -{ \gamma}^a{\widehat
\nabla}_\mu =0\label{nablagammaa}
\ee
The axial conjugate relation holds as well. Therefore
\be
{\widehat  \gamma}^\mu  \widehat\nabla_\mu \,\overline{\widehat
\gamma}^\nu\overline{ \widehat\nabla}_\nu=
\gamma^a \gamma^b \overline{\widehat e}_a^\mu  \overline{\widehat e}_b^\nu
\overline{\widehat \nabla}_\mu\overline{\widehat \nabla}_\nu= \eta^{ab}
\overline{\widehat e}_a^\mu  \overline{\widehat e}_b^\nu  \overline{\widehat
\nabla}_\mu\overline{\widehat \nabla}_\nu+ \Sigma^{ab}\overline{\widehat
e}_a^\mu  \overline{\widehat e}_b^\nu  [\overline{\widehat \nabla}_\mu,
\overline{\widehat \nabla}_\nu]\label{sym+antisym}
\ee

 On the other hand, when acting on a (bi-)spinor quantity
\be
\Sigma^{ab}\overline{\widehat
e}_a^\mu  \overline{\widehat e}_b^\nu  [\overline{\widehat \nabla}_\mu,
\overline{\widehat \nabla}_\nu] = \frac 18 \gamma^a
\gamma^b\gamma^c\gamma^d  \widehat R_{abcd}=-\frac 14 \widehat
R_{\mu\nu\lambda\rho} \widehat g^{\mu\lambda}\widehat g^{\nu\rho}=-\frac 14
\widehat R\label{gammaR}
\ee
where use is made of
\be
 \widehat R_{abcd}= \widehat e_a^\mu \widehat e_b^\nu \widehat e_c^\lambda
\widehat e_d^\rho \widehat R_{\mu\nu\lambda\rho}. \label{Ramu}
\ee

Now replacing \eqref{hatansatz1} into \eqref{hatFpropa} and using the above we
get
\be
\sqrt{|\overline{\widehat g}|}\left( \overline{\widehat\nabla}_\mu
\overline{\widehat g}^{\mu\nu}\overline {\widehat\nabla}_\nu -\frac 14\overline{
\widehat R}\right)
 \overline{\widehat \EG}(\widehat x,\widehat x')= -{\bf 1}{\mathbf{\delta
}}(\widehat x,\widehat x')\label{hatFpropa1}
\ee

The differential operator acting on $\overline{\widehat \EG}$ will be denoted	
by
$\overline{\widehat \EF}_{\hat g}$. In compact operator notation
\be
\overline {\widehat \EF}_{\hat g}\overline {\widehat \EG}_{\hat g}=-{\bf
1},\label{hatFG}
\ee
with $\langle \widehat x|\widehat \EG_{\hat g}| \widehat x'\rangle = 
\overline{\widehat
\EG}_{\hat g}( \widehat x, \widehat x')$.

As a consequence of \eqref{Fdagger} we have
\be
\left[\sqrt{\widehat g}\left(  {\widehat\nabla}_\mu
 {\widehat g}^{\mu\nu} {\widehat\nabla}_\nu -\frac 14 {
\widehat R}\right)\right]^\dagger=
\eta \left[\sqrt{| {\widehat g}|}\left(  {\widehat\nabla}_\mu  {\widehat
g}^{\mu\nu} {\widehat\nabla}_\nu -\frac 14 { \widehat
R}\right)\right]\eta\label{DDdagger}
\ee
or
\be
\left( {\widehat \EF}_{\hat g}\right)^\dagger =\eta\, \widehat \EF_{\hat
g}\,\eta\label{FgFg}
\ee
We shall refer often to the related operator
\be
\widehat\EF=\frac 1{ \sqrt{\widehat g}}\,\,\widehat \EF_{\hat g}, \quad\quad
{\widehat \EF}^\dagger=
\eta\, {\widehat \EF}\eta \label{hatEF}
\ee
and to its inverse $\widehat\EG$: $\widehat \EF \widehat \EG=-{\bf 1}$.

\vskip 0.5 cm
{\bf Remark 3.} The operator $\widehat \EF$ is the main intermediate
result of our paper. It is natural to assume that its inverse  $\widehat
\EG$ exists. There is no reason to believe that it does not, because,
the differential operator $\widehat \EF$ (after a Wick rotation) can be defined
as an 
axial-elliptic operator, at least under reasonable conditions  on  the axial
tensor $f_{\mu\nu}$. 
In fact its quadratic part can be cast in the form $-\partial_i  A_{ij}(x)  
\partial_j$, 
where $A_{ij}$ is an invertible matrix and its dominating part is symmetric and
positive definite.
However, no doubt, it would be desirable to  have a mathematical
(possibly constructive) proof of the existence of $\widehat \EG$ . In Appendix C
we discuss this
issue and, following \cite{DeWitt1}, we give some arguments in this direction.

\section{The Schwinger proper time method}

From now on, for practical reasons, we drop the bar symbol of axial conjugation.
At the end we will axially-conjugate the result.

Let us define  the amplitude
\be
\langle\widehat  x,\widehat s| \widehat x',0\rangle = \langle\widehat  x| e^{i 
\widehat\EF
\widehat s}|\widehat x'\rangle\label{hatampA}
\ee
which satisfies the (heat kernel) differential equation
\be
i \frac {\partial }{\partial \widehat s} \langle \widehat x,\widehat s|\widehat 
x',0\rangle
= - \widehat\EF_{\hat x} \langle
\widehat x,\widehat s|\widehat  x',0\rangle\equiv K(\widehat x, \widehat x',
\widehat s) \label{hatdiffeqforA}
\ee
where $\widehat\EF_{\hat x}$ is the differential operator
\be
\widehat\EF_{\widehat x}=\widehat\nabla_\mu {\widehat g}^{\mu\nu}
\widehat\nabla_\nu- \frac
14 \widehat R\label{hatEHx}
\ee
Then we make the ansatz
\be
\langle \widehat x,\widehat s| \widehat x',0\rangle = -\lim_{m\to 0}\frac
i{16\pi^2}
\frac {\sqrt{\widehat
D(\widehat x,\widehat x')}} {\widehat s^2}
e^{i\left(\frac {\widehat\sigma(\widehat x,\widehat x')} {2\widehat
s}-m^2\widehat s\right)}\widehat
\Phi(\widehat x,\widehat x',\widehat s)\label{ansatz3}
\ee
where $\widehat D(\widehat x,\widehat x')$ is the VVM determinant and
$\widehat\sigma$ is the world function
(see above). $\widehat\Phi(\widehat x,\widehat x',\widehat s)$ is a
function to be determined. It is useful to introduce also the mass parameter
$m$, which
we will eventually set to zero.
In the limit $\widehat  s\to 0$ the RHS of \eqref{ansatz3} becomes the
definition
of a delta function multiplied by $\widehat\Phi$. More precisely,  since it must
be $\langle\widehat x,0|\widehat  x',0\rangle=\delta(\widehat x,\widehat x')$,
and
\be
\lim_{\widehat s\to 0} \frac i{4\pi^2} \frac {\sqrt{\widehat D(\widehat
x,\widehat x')}} {\widehat s^2}
\, e^{i\left(\frac {\widehat\sigma(\widehat x,\widehat x')} {2\widehat
s}-m^2\widehat s\right)}
= \sqrt{|\widehat g(\widehat x)|}\,\,
\delta(\widehat x,\widehat x'),\label{hatlimdelta}
\ee
we must have
\be
\lim_{\widehat s\to 0}\widehat \Phi(\widehat x,\widehat x',\widehat s)={\bf
1}\label{hatlimitPhi}
\ee
Eq.\eqref{hatdiffeqforA} becomes an equation for $\widehat \Phi(\widehat
x,\widehat x',\widehat s)$. Using (\ref{smus}) and (\ref{DeltaTrace}), after
some algebra one gets
\be
i\frac {\partial \widehat\Phi}{\partial\widehat  s} +\frac i{\widehat s}
\widehat \nabla^\mu
\widehat \Phi \widehat \nabla_\mu\widehat \sigma
+\frac 1{\sqrt{\widehat D}} \widehat \nabla^\mu \widehat \nabla_\mu
\left(\sqrt{\widehat D} \widehat \Phi\right)-\left(\frac 14 \widehat
R-m^2\right)\widehat\Phi=0\label{hateqforPhi}
\ee
Now we expand
\be
\widehat\Phi(\widehat x,\widehat x',\widehat s)
= \sum_{n=0}^\infty \widehat a_n(\widehat x,\widehat x') (i\widehat s)^n
\label{hatPhiexp}
\ee
with the boundary condition $[\widehat a_0]=1$. The $\widehat a_n$ must satisfy
the recursive relations:
\be
(n+1)\widehat a_{n+1} +\widehat  \nabla^\mu \widehat a_{n+1} \widehat \nabla_\mu
\widehat \sigma - \frac 1{\sqrt{\widehat D}} \widehat \nabla^\mu \widehat
\nabla_\mu
\left(\sqrt{\widehat D}  \widehat a_n\right) +\left(\frac 14 \widehat R
-m^2\right)\widehat  a_n=0
\label{hatrecursive}
\ee
Using these relations and the coincidence results of section 3.3, 3.4 and 3.5,
it is possible to compute
each coefficient $a_n$ at the coincidence limit.

\subsection{Computing $\widehat a_n$}

In this subsection we wish to compute $ {[\widehat a_1]}$ and  $ {[\widehat
a_2]}$, which will be needed
later on.
We start from \eqref{hatrecursive} for $n=-1$.:
\be
\widehat \nabla^\mu \widehat a_0 \, \sigma_{;\mu} =0, \quad\quad {\rm
with}\quad\quad [\widehat a_0]={\bf 1}, \label{a0eq}
\ee
which implies that
\be
\widehat a_0(\widehat x,\widehat x')= \widehat I(\widehat x,\widehat
x').\label{a0I}
\ee
Replacing this inside \eqref{hatrecursive} for $n=0$ one gets
\be
\widehat a_1(\widehat x,\widehat x') +  \widehat \nabla^\mu  \widehat \sigma
\nabla_\mu \widehat
a_1(\widehat x,\widehat x') -  \frac 1{\sqrt{\widehat \Delta}} \widehat
\nabla^\mu \widehat
\nabla_\mu
\left(\sqrt{\widehat \Delta}\,  \widehat I(\widehat x,\widehat x')\right)
+\left(\frac 14 \widehat
R
-m^2\right)\widehat I(\widehat x,\widehat x') =0,\label{hata1eq}
\ee
which implies
\be
[\widehat a_1]= \left(-\frac 1{12} \widehat R + m^2 \right){\bf 1}\label{hata1}
\ee
Moreover differentiating \eqref{hata1eq} with respect to $\nabla_\lambda$ and
taking the coincidence limit:
\be
2[\widehat \nabla_\lambda \widehat a_1] &=& \frac 14 \widehat R_{;\lambda}{\bf
1} -[ \sqrt{\widehat \Delta}_{;\mu}{}^\mu{}_\lambda  \widehat I+
\widehat \nabla_\lambda \widehat\nabla^\mu \widehat \nabla_\mu\widehat I]\0
\ee
so
\be
[ \widehat\nabla_\lambda \widehat a_1]&=& 
{\left( 
\frac 1{12} \widehat{\cal R}_{\lambda\nu;}{}^\nu
-\frac 1{24} \widehat R_{;\lambda}\right)} 
{\bf 1}.\label{nablahata1}
\ee
Next we have
\be
[\widehat\nabla^\lambda \widehat\nabla_\lambda\left( \widehat a_1+
\widehat\nabla^\mu\widehat \sigma\, \widehat\nabla_\mu \widehat a_1\right)] =
3 [\widehat\nabla^\lambda \widehat\nabla_\lambda  \widehat a_1]\0
\ee
so that
\be
[\widehat\nabla^\lambda \widehat\nabla_\lambda  \widehat a_1]&=& \frac 13 [
\widehat\nabla^\lambda \widehat\nabla_\lambda \left(
 \frac 1{\sqrt{\widehat \Delta}} \widehat \nabla^\mu \widehat \nabla_\mu
\left(\sqrt{\widehat \Delta}\,  \widehat I\right) -\left(\frac 14 \widehat R
-m^2\right)\widehat I\right)]\label{nablanablaa1}\\
&=& \frac 13 \left( -\frac 1{20} \widehat R_{;\mu}{}^\mu
-\frac 1{30} \widehat R_{\mu\nu}\widehat R^{\mu\nu} + \frac 1{30} \widehat
R_{\mu\nu\lambda\rho}\widehat R^{\mu\nu\lambda\rho}+\frac 1{8}
\widehat{\cal R}_{\mu\nu}\widehat{\cal R}^{\mu\nu}\right)
\ee
Finally
\be
[\widehat a_2] &=&\frac 12 [ \widehat\nabla^\lambda \widehat\nabla_\lambda
\widehat a_1
- \left(\frac 1{12} \widehat R-m^2\right) \widehat a_1]\label{hata2}\\
&=& \frac 12 m^4 -\frac 1{12} m^2 \widehat R +\frac 1 {288} \widehat R^2 -\frac
1{120} \widehat R_{;\mu}{}^\mu
-\frac 1{180} \widehat R_{\mu\nu}\widehat R^{\mu\nu} + \frac 1{180} \widehat
R_{\mu\nu\lambda\rho}\widehat R^{\mu\nu\lambda\rho}+\frac  1{48}
\widehat{\cal R}_{\mu\nu}\widehat{\cal R}^{\mu\nu}   \0
\ee
We recall that $\widehat{\cal R}_{\mu\nu}= {\widehat R}_{\mu\nu}{}^{ab}
\Sigma_{ab}$.

\section{The odd trace anomaly}

We are now ready to compute that odd parity trace anomaly. Beside the
point-splitting,
which we have used above, we need a regulator to get rid of the infinities at
coincident point.
We will use two regularizations: the dimensional and zeta function ones.

\subsection{Schwinger-DeWitt and dimensional regularization}

We start again from the Dirac operator \eqref{hatF}.  We have defined above the
covariant square
\be
\widehat \EF =- \overline{\widehat F} {\widehat F}\label{EF}
\ee
We identify the effective action for Dirac fermions with
\be
\widehat W =-\frac i2 \Tr \left(\ln \,\widehat \EF\right)\label{EAEFg}
\ee
$\Tr$ includes also the spacetime integration.
The AE Weyl variation of \eqref{EAEFg} is given by
\be
\delta_{\widehat \omega }\widehat W = \frac i2 \Tr\left( \widehat \EG\,
\delta_{\widehat \omega }\widehat \EF\right)\label{deltaEA}
\ee
where
\be
\widehat\EF \widehat \EG =-1\label{EFginverse}
\ee

So we can write
\be
\delta_{\widehat \omega }\widehat W=\delta_{\widehat\omega} \left( -\frac 12
\int_0^\infty
\frac {d\widehat s}{i\widehat s}
e^{i\widehat \EF \,\widehat s}\right) = -\frac 12 \Tr\left( \int_0^\infty
d\widehat s \,
e^{i\widehat \EF \,\widehat s}\delta_{\widehat \omega}\widehat \EF\right).
\label{deltaEAW}
\ee
It follows that, as far as the variation with respect to axial-Weyl transform is
concerned,
the effective action can be represented as
\be
\widehat W= -\frac 12 \int_0^\infty \frac {d\widehat s}{i\widehat s}
e^{i\widehat \EF\widehat s}+ {\rm const}\equiv
\widehat  L+{\rm const}\label{W+const}
\ee
where $\widehat L$ is the relevant  effective action
\be
\widehat L=\int d^d\widehat x \, \widehat L(\widehat x) \label{Llocal}
\ee
which can be written  as
\be
\widehat L(\widehat x) =  -\frac 12 \tr \int_0^\infty \frac {d\widehat
s}{i\widehat s}
\widehat K(\widehat x,\widehat x',\widehat s)\label{WK}
\ee
where the kernel $\widehat K$ is defined by
\be
\widehat K(\widehat x,\widehat x ',\widehat s)= e^{i\widehat \EF \,\widehat s}
\delta(\widehat x,\widehat x')\label{Kxx'}
\ee
Inserted in $\delta_{\hat\omega}\widehat W$, under the symbol $\Tr$, it means
integrating over $x$ after
taking the limit $x'\to x$. So, looking at \eqref{ansatz3}, in dimension $d$,
\be
 \widehat K(\widehat x,\widehat x,\widehat s)
=\frac i{(4\pi i \widehat s)^{\frac d2}}\,\sqrt{\widehat g}\, e^{-im^2\widehat
s} [\widehat
\Phi(\widehat x,\widehat x,\widehat s)]
\label{Kxx}
\ee

A specification is in order at this point. For the heat kernel method to work a
Riemannian metric is required. Therefore at this stage we Wick-rotate the
metric, so that the operator $\widehat{\EF}$ becomes axial-elliptic. This
operation is understood from now on. After calculating the anomaly we will
return to the Lorentz signature.

\subsection{Analytic continuation in $d$}

The purpose now is to analytically continue in $d$. But we can do this only for
dimensionless quantities. We therefore multiply $\widehat L$ by $\mu^{-d}$,
where $\mu$
is a mass parameter. We have for a Dirac fermion
\be
\frac {\widehat L(x)}{\mu^d}= -\frac i2 (4\pi \mu^2) \tr
\int_0^{\infty}d\widehat s\,  (4\pi i
\mu^2 \widehat s)^{-\frac d2-1}\sqrt{\widehat g}e^{-im^2\widehat s} [\widehat
\Phi(\widehat x,\widehat x,\widehat s)]\label{Lxmud}
\ee
where $\tr$ denotes the trace over gamma matrices.

Now we make the assumption that
\be
\lim_{s\to\infty} e^{-im^2\widehat s} [\widehat \Phi(\widehat x,\widehat
x,\widehat s)]=0\label{assumption}
\ee
As a consequence we can integrate by parts
\be
\frac {\widehat L(x)}{\mu^d}&=&\frac i{d} \tr \int_0^{\infty}d\widehat s \frac
{\partial}{\partial (i\widehat s)}  (4\pi i \mu^2 \widehat s)^{-\frac
d2}\sqrt{\widehat
g}e^{-im^2\widehat s} [\widehat \Phi(\widehat x,\widehat x,\widehat
s)]\label{Lxmud1}\\
&=& - \frac i{d} \tr \int_0^{\infty}d\widehat s\, (4\pi i \mu^2 \widehat
s)^{-\frac
d2}\sqrt{\widehat g}\frac {\partial}{\partial(i\widehat s)} \left(
e^{-im^2\widehat s} [\widehat
\Phi(\widehat x,\widehat x,\widehat s)]\right) \0\\
&=&  \frac{2 i}{d(2-d)4\pi \mu^2} \tr \int_0^{\infty}d\widehat s\,  (4\pi i
\mu^2
\widehat s)^{1-\frac d2}\sqrt{\widehat g}\frac {\partial^2}{\partial(i\widehat
s)^2} \left(
e^{-im^2\widehat s} [\widehat \Phi(\widehat x,\widehat x,\widehat s)]\right)\0\\
&=& - \frac{4i}{d(2-d)(4-d)}\frac 1{(4\pi \mu^2)^2} \tr \int_0^{\infty}d\widehat
s\,
(4\pi i \mu^2 \widehat s)^{2-\frac d2}\sqrt{\widehat g}\frac
{\partial^3}{\partial(i\widehat s)^3}
\left( e^{-im^2\widehat s} [\widehat \Phi(\widehat x,\widehat x,\widehat
s)]\right)\0
\ee
Next we use
\be
[\widehat \Phi(\widehat x,\widehat x,\widehat s)]= 1+ {[\widehat a_1]} i\widehat
s+ {[\widehat a_2]}(i\widehat s)^2+\ldots\label{Phiexpansion}
\ee
and, around $d=2$, we use $\frac 1{d(2-d)} = \frac 12 \left( \frac 1{d-2}- \frac
1d\right)$ and in the third line of \eqref{Lxmud1} we use
\be
(4\pi i \mu^2 s)^{1-\frac d2}= 1- \frac {d-2}2 \ln(4\pi i \mu^2 s)+\ldots\0
\ee
Then we differentiate once $ [\widehat \Phi(\widehat x,\widehat x,\widehat s)]$,
and the remaining
derivation we get rid of
by integrating by parts. Finally one gets
\be
\widehat L(\widehat x)&=& \frac 1{4\pi} \left(\frac 1{d-2}-\frac 12\right) \tr
\left(
({[\widehat a_1]}-m^2)\sqrt{\widehat g}\right)  \label{Lxd2}\\
&& -\frac i{8 \pi}\tr \int_0^{\infty}d\widehat s\,  \ln(4\pi i \mu^2 \widehat
s)\sqrt{\widehat
g}\frac {\partial^2}{\partial(i\widehat s)^2} \left( e^{-im^2\widehat s}
[\widehat
\Phi(\widehat x,\widehat x,\widehat s)]\right)\0
\ee

Around $d=4$ we use $\frac 1{d(d-2)(d-4)}\approx \frac 18 \left( \frac 1{d-4}
-\frac 34\right)$. With reference to the last line of \eqref{Lxmud1}, we
differentiate twice $ [\widehat \Phi(x,x,s)]$ and integrate by parts the third
derivative. The result is
\be
\widehat L(\widehat x) &\approx& \frac 1{32\pi^2}\left( \frac 1{d-4} -\frac
34\right)\tr
\left(m^4-2m^2 {[\widehat a_1]}+2 {[\widehat a_2]}\right)\sqrt{\widehat
g}\label{Lxd4}\\
&&+  \frac i{64\pi^2}\tr \int_0^{\infty}d\widehat s\,  \ln(4\pi i \mu^2 \widehat
s)
\sqrt{\widehat g}\frac {\partial^3}{\partial(i\widehat s)^3} \left(
e^{-im^2\widehat s} [\widehat
\Phi(\widehat x,\widehat x,\widehat s)]\right)\0
\ee
The last line depends explicitly on the parameter $\mu$ and represent a nonlocal
part

\subsection{The anomaly}

Let us take the variation of \eqref{Lxd4} with respect to $\widehat \omega=
\omega+\gamma_5 \eta$.  

Recall that
\be
\delta_{\widehat\omega} \sqrt{\widehat g}&=& d \,\widehat\omega\, \sqrt{\widehat
g}\label{deltaomega1}\\
\delta_{\widehat\omega} \widehat R&=& -2\widehat \omega \,\widehat R -2 (d-1)
\widehat \square \widehat \omega\label{deltaomega2}\\
\delta_{\widehat\omega} \widehat R_{\mu\nu\lambda}{}^\rho &=&-\delta_\nu^\rho
\widehat D_\mu \widehat D_\lambda \widehat \omega +  \delta_\mu^\rho \widehat
D_\nu \widehat D_\lambda \widehat \omega+ \widehat D_\mu \widehat D_\sigma
\widehat \omega \,\widehat g^{\rho\sigma} \widehat g_{\nu\lambda}-\widehat D_\nu
\widehat D_\sigma \widehat \omega\, \widehat g^{\rho\sigma} \widehat
g_{\mu\lambda}\label{deltaomega3}
\ee
From these follows, for instance,
\be
\delta_{\widehat\omega} \left(\sqrt{\widehat g}\widehat R^2\right) = (d-4)
\sqrt{\widehat g}\,\widehat \omega\,\widehat R^2- 4(d-1) \widehat R\,
\sqrt{\widehat g}\, \widehat \square \widehat \omega\label{deltaomega4}
\ee
{
\be
\delta_{\widehat\omega} \left(\sqrt{\widehat g}\widehat
R_{\mu\nu}\widehat R^{\mu\nu}\right)
&=& (d-4) \widehat \omega\, \sqrt{\widehat g}\,\widehat
R_{\mu\nu}\widehat R^{\mu\nu}
 +2(2-d) \sqrt{\widehat g}\, \widehat R^{\mu\nu} 
{\widehat D_\mu \widehat D_\nu} \widehat \omega
 -2 \sqrt{\widehat g}\, \widehat R
\widehat\square \widehat\omega
\0\\
&=& (d-4) \widehat \omega\, \sqrt{\widehat g}\,\widehat
R_{\mu\nu}\widehat R^{\mu\nu}
 -d \sqrt{\widehat g}\, \widehat R
\widehat\square \widehat\omega
\label{deltaomega5}
\ee}
\be
\delta_{\widehat\omega} \left(\sqrt{\widehat g}\widehat
R_{\mu\nu\lambda\rho}\widehat R^{\mu\nu\lambda\rho}\right)
&=& (d-4) \widehat \omega\, \sqrt{\widehat g}\,\widehat
R_{\mu\nu\lambda\rho}\widehat R^{\mu\nu\lambda\rho}
 -8 \sqrt{\widehat g}\, \widehat R^{\mu\nu} 
{\widehat D_\mu \widehat D_\nu} \widehat \omega
\0\\
&=& {(d-4) \widehat \omega\, \sqrt{\widehat g}\,\widehat
R_{\mu\nu\lambda\rho}\widehat R^{\mu\nu\lambda\rho}
 -4 \sqrt{\widehat g}\, \widehat R
\widehat\square \widehat\omega}
\label{deltaomega6}
\ee
\be
\delta_{\widehat\omega} \left(\sqrt{\widehat g}
\widehat \square \widehat R\right)
&=& (d-4) \widehat \omega\, \sqrt{\widehat g}\,\widehat \square \widehat R
+(d-6) \sqrt{\widehat g}\, \partial_\mu \widehat \omega\, \partial^\mu \widehat
R
 -2 \sqrt{\widehat g}\, \widehat R \, \widehat \square \, \widehat \omega
 -2 (d-1) \sqrt{\widehat g} \, \widehat \square^2 \, \widehat \omega
 \0\\
&=& 0\0
\ee
and
\be
\delta_{\widehat\omega} 
\tr\left( 
\sqrt{\widehat g}\,\widehat {\cal R}_{\mu\nu}\widehat{\cal R}^{\mu\nu} \right) 
&=& (d-4)  \tr\left(\widehat\omega \,
\sqrt{\widehat g}\,\widehat {\cal R}_{\mu\nu}\widehat{\cal R}^{\mu\nu} \right)
+ {4} \, \tr \left( \sqrt{\widehat g} \, \widehat R^{\mu\nu} 
{\widehat D_\mu \widehat D_\nu} \widehat\omega \right) 
\0\\
&=&   {(d-4) \tr\left(\widehat\omega \,
\sqrt{\widehat g}\,\widehat {\cal R}_{\mu\nu}\widehat{\cal R}^{\mu\nu} \right)
+2 \, \tr \left( \sqrt{\widehat g} \, \widehat R \widehat \square \widehat\omega
\right) }
\label{deltaomega7}
\ee

In the first line of \eqref{Lxd4} one can ignore $m^2$ or $m^4$ terms
(either one sets $m=0$ or they can be subtracted
because they are trivial). 
The second line \eqref{Lxd4} does not contain
singularities when $d\to 4$: it contains either vanishing or finite terms in
this limit. 
Let us denote the second line by $\widehat L_R$.
\be
\widehat L &=& \frac 1{16\pi^2}\left( \frac 1{d-4} -\frac
34\right)
\int d^d\widehat x\,
\tr
\left({[\widehat a_2]|_{m=0}}\sqrt{\widehat g}\right)\label{LR} + \widehat L_R
\ee
We now act with 
$\delta_{\widehat\omega} = \int d^d\widehat x \,
2 \tr \left(\widehat\omega\,\widehat g_{\mu\nu} {\frac{\delta}{\delta {\widehat
g_{\mu\nu}}}}\right)$%
\footnote{In MAT case, $\widehat g_{\mu\nu}$ also has two spinor indices, so
that 
$\omega\, g_{\mu\nu} {\frac{\delta}{\delta {g_{\mu\nu}}}} 
\rightarrow \widehat\omega_{AB}\,\widehat g_{\mu\nu}{}_{BC}
{\frac{\delta}{\delta {\widehat g_{\mu\nu}{}_{AC}}}}$. 
Since in our case $\gamma^5$ is symmetric, we have $\widehat a_{AB} = \widehat
a_{BA}$ and we can write $\delta_{\widehat \omega}$ as 
$\int d^d\widehat x \,
2 \tr \left(\widehat\omega\,\widehat g_{\mu\nu}{} {\frac{\delta}{\delta
{\widehat g_{\mu\nu}{}}}}\right)$. }

.
From \eqref{deltaomega1}-\eqref{deltaomega5} it follows that
\be
\delta_{\widehat\omega} \tr \left(\sqrt{\widehat g} \,[\widehat
a_2]|_{m=0}\right) 
= (d-4)\tr \left( \sqrt{\widehat g} \, \widehat\omega \, [\widehat a_2]|_{m=0} 
\right) 
-\frac{d-4}{120}\tr \left( \sqrt{\widehat g} \, \widehat R \widehat \square
\widehat\omega \right) 
\label{totald4}
\ee
The second piece can be canceled e.g.\ by a counterterm proportional to 
$\tr \left( \sqrt{\widehat g}\widehat R^2 \right)$.
Using the fact that the bare part of the action is Weyl invariant 
$\delta_{\widehat \omega} \widehat L = 0$ 
and that the renormalised part $\widehat L_R$ defines the (quantum) energy
momentum tensor
$\frac{2}{\sqrt{\widehat g}}\frac{\delta}{\delta {\widehat g_{\mu\nu}}} \widehat
L_R = \widehat\Theta^{\mu\nu}$ we get
\be
\int d^d\widehat x \, \tr \left(\widehat\omega {\sqrt{\widehat g}} \, \widehat
g_{\mu\nu} \widehat\Theta^{\mu\nu}\right)
&=& -\frac 1{16\pi^2}\int d^d\widehat x \tr \left( \sqrt{\widehat g} \,
\widehat\omega\, [\widehat a_2]|_{m=0} \right)  \label{LR2}
\ee
where
the $d-4$ factor in \eqref{totald4} canceled
the pole $\frac 1{d-4}$ in \eqref{LR}.

Clearly, the odd parity anomaly
can come only from the term
$\widehat{\cal R}_{\mu\nu}\widehat{\cal R}^{\mu\nu}$ 
contained in
$ {[\widehat a_2]}$  , with a coefficient of $\frac
1{32\pi^2}$ (for Majorana fermions, $\times 2$ for Dirac fermions). 
For the odd part we have
\be
\int d^d\widehat x  \, \tr {\sqrt{\widehat g}}\,\widehat\omega \, \widehat\ET
= -\frac 1{768 \pi^2}  \int d^4x \,\tr\sqrt{\widehat g}\, \widehat\omega
\,\widehat{\cal R}_{\mu\nu}\widehat{\cal R}^{\mu\nu}\Big{\vert}_{\rm odd}
\label{intan}
\ee
where we denoted $\widehat \ET =  \widehat g_{\mu\nu}  \widehat \Theta^{\mu\nu}
=  \widehat g_{\mu\nu} \langle\!\langle \widehat T^{\mu\nu}\rangle\!\rangle$.

The (odd parity) coefficient of $\omega$ defines 
{$\ET$}
and the (odd
parity) coefficient of $\eta$ defines 
{$\ET_5$}.
{Setting $\widehat\ET = \ET + \gamma_5 \ET_5$}
one obtains in this way
\be
{\ET} = -{\frac 14}\frac {1}{768\pi^2} \tr
\left({\widehat
{\cal R}}_{\mu\nu}\widehat{\cal R}^{\mu\nu} \right)\Big{\vert}_{\rm odd}=
- {\frac 14}\frac {2i}{768\pi^2} \epsilon^{\mu\nu\lambda
\rho}R^{(1)}_{\mu\nu\alpha\beta}
R^{(2)}_{\lambda\rho}{}^{\alpha\beta} \label{traceanomThetamumu}
\ee
and
\be
{\ET_5} = -{\frac 14}\frac {1}{768\pi^2} \tr
\left(\gamma_5\widehat{\cal R}_{\mu\nu}\widehat{\cal R}^{\mu\nu}
\right)\Big{\vert}_{\rm odd} =-{\frac 14}
\frac {i}{768\pi^2} \epsilon^{\mu\nu\lambda
\rho}\left(R^{(1)}_{\mu\nu\alpha\beta} R^{(1)}_{\lambda\rho}{}^{\alpha\beta}
+ R^{(2)}_{\mu\nu\alpha\beta} R^{(2)}_{\lambda\rho}{}^{\alpha\beta}\right)
\label{traceanomTheta5mumu}
\ee
In the last step we have Wick-rotated back the result: this is the origin of the
$i$ in the anomaly coefficient.
At this point we can safely set $x_2^\mu=0$ everywhere.

\subsection{$\zeta$-function regularization}

Given a differential operator $A$  in analogy with the Riemann $\zeta$ function,
the expression
$A^{-z}$, for complex $z$, is called $\zeta$ function regularization of $A$:
\be
\zeta(z,A)= A^{-z}= \frac 1{\Gamma(z)} \int_0^\infty dt \,t^{z-1} \,
e^{-tA}\label{A-z}
\ee
We will apply this representation to the operator $\widehat \EF(\widehat
x,\widehat x)$, :
\be
 (\widehat \EF(\widehat x ))^{-z}= \frac 1{\Gamma(z)} \int_0^\infty dt \,t^{z-1}
\,
\langle \widehat x|e^{-t \widehat \EF}|\widehat x\rangle\label{EF-z}
\ee
where $ \langle\widehat  x|e^{-t \widehat \EF}|\widehat x\rangle$ means the
coincidence limit of
$ \langle \widehat x|e^{-t \widehat \EF}|\widehat x'\rangle$. Eq.\eqref{EF-z} is
not quite
correct because only dimensionless quantities can be raised to an arbitrary
power. Moreover the object of interest will be $\widehat\EG$, rather than
$\widehat \EF$. Thus we introduce again the mass parameter $\mu$ and shift
from $t$ to $i\widehat s\mu$.
\be
\zeta(\widehat x,z)\equiv(\mu^2\widehat \EG(\widehat x,\widehat x))^{z}= \frac
1{\Gamma(z)} \int_0^\infty
(i\mu^2)d\widehat s \,(i\widehat s \mu^2)^{z-1} \, \langle x|e^{i \widehat s 
\widehat
\EF}|\widehat x\rangle\label{mu2EG-z}
\ee
Finally we replace $\langle\widehat  x|e^{i\widehat  s  \widehat \EF}|\widehat
x\rangle$ with $\widehat K(\widehat x,\widehat x,\widehat s)$
in eq.\eqref{Kxx}. The result is
\be
\zeta(\widehat x,z)=(\mu^2\widehat \EG(\widehat x,\widehat x))^{z}= \frac
i{\Gamma(z)} \mu^d
\sqrt{\widehat g} \int_0^\infty (i\mu^2)d\widehat s \,(i\widehat s
\mu^2)^{z-1-\frac d2}
\,e^{-im^2\widehat s} [\widehat \Phi(\widehat x,\widehat x,\widehat s)] 
\label{mu2EG-zfin}
\ee
which can be rewritten as
\be
 \zeta(\widehat x,z)=(\mu^2\widehat \EG(\widehat x,\widehat x))^{z}&=& -\frac
i{\Gamma(z)}
\frac{\mu^{d-4}}{(4\pi)^{\frac d2}} \frac{\sqrt{\widehat g}}{(z-\frac
d2)(z-\frac d2 +1)(z-\frac d2+2)}\0\\
&& \times \int_0^\infty d(i\widehat s)\,(i\widehat s \mu^2)^{z-\frac d2+2}\frac
{\partial^3}
{ \partial(i\widehat s)^3} \left(e^{-im^2\widehat s} [\widehat \Phi(\widehat
x,\widehat x,\widehat s)]\right)
\label{mu2EG-zfin1}
\ee
This is well defined for $d=4$ at $z=0$.
\be
\zeta(\widehat x,0)= \frac {i\sqrt{\widehat g}}{2 (4\pi)^2}\left[ \frac
{\partial^2}
{ \partial(i\widehat s)^2} \left(e^{-im^2\widehat s} [\widehat
\Phi(\widehat x,\widehat x,\widehat s)]\right)\right]_{\widehat
s=0}\label{zetax0}
\ee

Now, differentiating \eqref{A-z} with respect to $z$ and evaluating at $z=0$, we
get formally
\be
\frac d{dz}\zeta(z,A)\vert_{z=0} = - \Tr \ln A\label{dzetaz}
\ee
This suggest the procedure to regularize $\widehat W$ (which is  the trace of a
log). More precisely
\be
\widehat W\rightarrow \widehat W_\zeta = - \frac i2 \zeta'(0), \quad\quad {\rm
where}\quad\quad
\zeta(z)= \int \tr\, \zeta(\widehat x,z) d^d\widehat x\label{Wzeta}
\ee
As a consequence for $d=4$:
\be
\widehat L_\zeta(x) &=&\frac 1{64\pi^2}(\gamma+\frac 32)  \sqrt{\widehat g} \,
\tr
\left(2[\widehat a_2\widehat (x)] -2m^2 [\widehat a_1(\widehat x)]
+m^4\right)\label{Lzeta}\\
&& -\frac i{64\pi^2}   \sqrt{\widehat g}\int_0^\infty d\widehat s\, \ln(4\pi i
\mu^2 \widehat s)
\frac {\partial^3}
{ \partial(i\widehat s)^3} \left(e^{-im^2\widehat s} [\widehat \Phi(\widehat
x,\widehat x,\widehat s)]\right)\0
\ee

Now, suppose that the operator $A$, under a symmetry transformation with
parameter $\epsilon$,
transforms  as
\be
\delta_\epsilon A = \{A,\epsilon\}.\label{deltaA}
\ee
Then
\be
\delta_\epsilon\Tr  A^{-z} = -2z \Tr \left(A^{-z}\epsilon\right)  = -2z \Tr
\left( \zeta(z,A) \epsilon\right)\label{deltatrA}
\ee
Since the relevant result is obtained by differentiating with respect to $z$ and
setting $z=0$, once the functional is regularized, the anomalous part of the
effective action is extremely easy to derive:
\be
 \widehat L_{\cal A}=- 2\Tr  \left( \zeta(0,A) \epsilon\right)\label{Lanomalous}
\ee

Let us return to the our problem. The operator to be regulated is 
${\widehat \EF}=  \widehat \EF_{\hat x}$. Its AE Weyl transformation is
{\be
\delta_{\hat \omega} {\widehat \EF} &=& -2 \widehat \omega \, {\widehat \EF} 
+ \left(
{\overline{\widehat\gamma}}{}^\mu 
\widehat \gamma^\nu 
 +  \widehat g^{\mu\nu}\right)
 \partial_\nu \widehat\omega \widehat \nabla_\mu +\frac 32
\widehat \square \widehat \omega \0\\
&=&-2 \widehat \omega \, {\widehat \EF}+    {\widehat \EF}\left[
\frac 1{\widehat \EF}\left( \left(
{\overline{\widehat\gamma}}{}^\mu 
\widehat \gamma^\nu 
 +  \widehat g^{\mu\nu}\right)
 \partial_\nu \widehat\omega \widehat \nabla_\mu +\frac 32
\widehat \square \widehat \omega\right)\right]\0
\ee
}
$\widehat \EG(\widehat x,\widehat x)$ is the inverse of $\widehat \EF$ and its
transformation is similar:
{ \be  
\delta_{\hat \omega} \widehat \EG= 2  \,{\widehat \EG}
\,\widehat \omega+   
{\widehat \EG}\left[
\left( \left(
{\overline{\widehat\gamma}}{}^\mu 
\widehat \gamma^\nu 
 +  \widehat g^{\mu\nu}\right)
 \partial_\nu \widehat\omega \widehat \nabla_\mu +\frac 32
\widehat \square \widehat \omega\right)
 {\widehat\EG}\right]\0
\ee}
The first piece in the RHS reproduces exactly the mechanism in \eqref{deltatrA}.
The second is a nonlocal term of the effective action; it does not concern us
here and we drop it.
As noticed above this procedure does not lead directly to the anomaly. It rather
gives the anomalous part of the effective action, i.e. the anomaly integrated
with the insertion of ${\sqrt{\widehat g}}$:
\be
\widehat L_{\cal A}(\widehat \omega) &=& -i \Tr \left( \widehat \omega \,
\zeta(\widehat x,0)\right)\label{anomhatomega}\\
&=&i\, \Tr \left(\frac {\sqrt{\widehat g}}{2 (4\pi)^2}\left[ \frac {\partial^2}
{ \partial(i\widehat s)^2} \left(e^{-im^2\widehat s} [\widehat \Phi(\widehat
x,\widehat x,\widehat s)]\right)\right]_{s=0}
\widehat \omega \right)\0\\
&=&i\,\Tr \left(\frac {\sqrt{\widehat g}}{2 (4\pi)^2} \left(2[\widehat
a_2(\widehat x)] -2m^2
[\widehat a_1(\widehat x)] +m^4\right)\,\widehat \omega \right)\0
\ee
Now, proceeding as before, we differentiate with respect to $\widehat \omega$
and strip off ${\sqrt{\widehat g}}$, multiply back $\widehat\omega$ and obtain
the true integrated anomaly. This leads to the same results as above.

\subsection{The collapsing limit}

After computing the trace anomalies \eqref{traceanomThetamumu} and
\eqref{traceanomTheta5mumu} of a Dirac fermion coupled to a metric and an axial
symmetric tensor, we are now interested in returning to the original problem,
that is the trace anomaly of a Weyl tensor in an chiral fermion theory coupled
to ordinary gravity. To this end we take the collapsing limit. { In \cite{MAT1}
the latter was defined as $h_{\mu\nu}\to \frac
{h_{\mu\nu}}2,k_{\mu\nu} \to \frac {h_{\mu\nu}}2$, with $h_{\mu\nu}$ and
$k_{\mu\nu}$ both infinitesimal.
Here we do not put such a limitation. The collapsing limit is defined by making
the replacements 
\be \label{colllr}
g_{\mu\nu} \to \eta_{\mu\nu} + \frac
{h_{\mu\nu}}2 \qquad,\qquad f_{\mu\nu} \to \frac {h_{\mu\nu}}2
\ee
in the previous formulas, with finite $h_{\mu\nu}$. With this choice one has
\be
\hat{g}_{\mu\nu} = \frac{1}{2} (1-\gamma^5)\, \eta_{\mu\nu} + \frac{1}{2}
(1+\gamma^5)\, G_{\mu\nu}
\qquad,\qquad G_{\mu\nu} \equiv \eta_{\mu\nu} + h_{\mu\nu}\label{gG}
\ee
From this we see that the left-handed part couples to the flat metric, 
while the right-handed part couples to the (generic) metric $G_{\mu\nu}$. As a
consequence we have also
\be
\widehat e^a_m \to \delta^a_m\frac {1-\gamma_5}2 + e^a_m\, \frac
{1+\gamma_5}2,\quad\quad
\widehat e^m_a \to  \delta_a^m \frac {1-\gamma_5}2 + e^m_a \, \frac
{1+\gamma_5}2, \label{vierbeinlimit}
\ee
as well as
\be
\sqrt{\widehat g}\rightarrow \frac {1-\gamma_5}2 + \frac {1+\gamma_5}2
\sqrt{G},\label{detGlimit}
\ee
Similarly for the Christoffel symbols
\be
 \Gamma_{\mu\nu}^{(1)\lambda} \to \frac 12
\Gamma_{\mu\nu}^\lambda,\quad\quad\Gamma_{\mu\nu}^{(2)\lambda} \to \frac 12
\Gamma_{\mu\nu}^\lambda,\label{Gamma12limit}
\ee
for the spin connections
\be
 \Omega_\mu^{(1)ab}\to \frac 12  \omega_\mu^{ab},\quad\quad 
\Omega_\mu^{(2)ab}\to \frac 12  \omega_\mu^{ab},\label{Omega12limit}
\ee 
and for the curvatures
\be
 R^{(1)}_{\mu\nu\lambda}{}^\rho\to \frac 12 R_{\mu\nu\lambda}{}^\rho,
\quad\quad
R^{(2)}_{\mu\nu\lambda}{}^\rho\to \frac 12
R_{\mu\nu\lambda}{}^\rho,\label{R12limit}
\ee
where all the quantities on the RHS of these limits are built with the metric
$G_{\mu\nu}$.}

As a consequence, the action \eqref{axialaction} becomes 
\be
\widehat S \longrightarrow S'=  \int d^4x \,  \left[ i\overline {\psi}\gamma^a
\frac{1-\gamma_5}2\partial_a \psi+ \int d^4x\, \sqrt{G}\, i\overline
{\psi}\gamma^a e_a^\mu
\left(\partial_\mu +\frac 12 \omega_\mu \right)\frac{1+\gamma_5}2\psi\right]  
\label{axialactionlimit}
\ee
where $\gamma^a$ is the flat (non-dynamical) gamma matrix while the vierbein
$e_a^\mu$ 
and the connection 
$\omega_\mu$ are compatible with the metric $G_{\mu\nu}$. Up to the term that
represents a decoupled left-handed fermion in the flat spacetime, the action
$S'$ 
is the action of a
right-handed Weyl fermion coupled to the ordinary gravity.

In the collapsing limit we have
\be
\ET(x) = \ET_5(x) =
- \frac{1}{16}\frac {2i}{768\pi^2} \epsilon^{\mu\nu\lambda
\rho}R_{\mu\nu\alpha\beta}
R_{\lambda\rho}{}^{\alpha\beta}
\ee
The integrated anomaly \eqref{intan} corresponding to $\widehat S$ thus becomes
\be
\int d^d\widehat x  \, \tr {\sqrt{\widehat g}}\,\widehat\omega \, \widehat\ET
&=& \int d^dx  \,  {\sqrt{G}}\,\left(\omega+\eta\right) \,
\left(\ET+\ET_5\right)\tr  P_+
+\int d^dx  \, \left(\omega-\eta\right) \, \left(\ET-\ET_5\right)\tr  P_-\0\\
&=&  4\int d^dx  \,  {\sqrt{G}}\,\omega_+ \, \ET\label{intanax}
\ee
where we used $\tr P_+=2$, $\ET-\ET_5=0$ and set $\omega_+= \omega +\eta$.
Notice that due to \eqref{gG} the transformation property of $G_{\mu\nu}$ is
$ G_{\mu\nu} \rightarrow e^{2\omega_+} G_{\mu\nu}$.
To extract an anomaly of the right fermion of the effective action corresponding
to \eqref{axialactionlimit} we take its Weyl variation with respect to the
metric $G_{\mu\nu}$
\be
\int d^dx  \,  {\sqrt{G}}\,\omega_+ \, \ET'\label{intanR}
\ee
where we denoted $\ET'=G_{\mu\nu}\Theta^{'\mu\nu}=G_{\mu\nu}\langle\!\langle 
T^{'\mu\nu}\rangle\!\rangle$.

Comparing \eqref{intanax} and \eqref{intanR}  we get
\be
\ET'(x) = - \frac {i}{1536\pi^2} \epsilon^{\mu\nu\lambda
\rho}R_{\mu\nu\alpha\beta}
R_{\lambda\rho}{}^{\alpha\beta}
\ee

If we instead of (\ref{colllr}) take the following collapsing limit
\be
g_{\mu\nu} \to \eta_{\mu\nu} + \frac
{h_{\mu\nu}}2 \qquad,\qquad f_{\mu\nu} \to -\frac {h_{\mu\nu}}2
\ee
then one obtains
\be
\hat{g}_{\mu\nu} = \frac{1}{2} (1-\gamma^5)\, G_{\mu\nu} + \frac{1}{2}
(1+\gamma^5)\, \eta_{\mu\nu}
\qquad,\qquad G_{\mu\nu} \equiv \eta_{\mu\nu} + h_{\mu\nu}
\ee
Now the right handed part is coupled to the flat metric and left handed part to
generic 
curved metric. We can now repeat the arguments from above and obtain the
Pontryagin 
Weyl anomaly for left-hended Weyl fermion
\be
\ET'(x) = \frac {i}{1536\pi^2} \epsilon^{\mu\nu\lambda
\rho}R_{\mu\nu\alpha\beta}
R_{\lambda\rho}{}^{\alpha\beta}.
\ee
The relative minus sign with respect to right-handed case is because of the
opposite 
sign in front of $\gamma_5$ matrix in the defining relation for projectors
$P_\pm$.

\section{Conclusion}

In \cite{BGL} the odd parity (Pontryagin) trace anomaly was calculated using a
Feynman diagram approach coupled to dimensional regularization. Only the lowest
order diagrams were computed, they allowed to identify the lowest order term of
the anomaly. The full anomaly was then reconstructed by covariantization, which
is correct if the diffeomorphisms are preserved by the regularization procedure.
This turned out to be the case, as was shown in \cite{BDL}.
After these two papers a negative result was obtained in \cite{Bast}. Using a
heat kernel  
method with a Pauli-Villars regularization the authors found a vanishing odd
parity trace anomaly in 4d. At this point it was imperative to find the culprit.
In \cite{BGL,BDL} the approach may appear too simple-minded, because only two
Feynman three-legged diagrams were considered, the triangle and the bubble
diagram. As was shown in the first part of \cite{MAT1} there are several
additional diagrams that may affect the final result. But, in fact, the accurate
analysis carried out in \cite{MAT1} showed that such additional diagrams cannot
change the result as far as the odd parity trace anomaly is concerned. It must
be admitted however that for such a delicate calculation an approach based
solely on Feynman diagrams may not be satisfactory. The  reason is the
preservation of chirality throughout the anomaly computation.

 It may appear obvious that if one wants to compute the anomaly of a left-handed
fermion coupled to gravity one has to respect its left-handedness and avoid
mixing different chiralities in the course of the computation. But this is not
as easy to do as to claim. As pointed out many times, the trouble arises with
the path integral measure, which is hard if not impossible to define for Weyl
fermions. If one uses a Fujikawa or heat kernel method (they are relatives) the
problem is transferred to the `square' of the Dirac operator, that is an
(Euclidean) elliptic operator that is used in these methods to define the
fermion determinant. The problem is: is there a quadratic operator that preserve
the same handedness as the linear Weyl operator? As was pointed out in
ref.\cite{MAT1} one such operator could be $\slashed{D}_L^\dagger
\slashed{D}_L$, where $ \slashed{D}_L= \slashed{D} P_L$ with $ \slashed{D}$ the
ordinary Dirac operator and $P_L$ the chiral projector, but, with this choice, a
phase would remain completely undetermined. We do not know if it is possible to
solve this problem, but we are sure the solution is not the choice made in
\cite{Bast}, because the operator chosen by the authors there includes both
chiralities. Of course, with this choice, the result for the odd trace anomaly
cannot be but 0.

A way out is provided by Bardeen's method, which we have used in this paper.
This method bypasses the difficulty mentioned above  because it utilizes Dirac
fermions, and so it is not hard to define a `square' Dirac operator, $\widehat
\EF$ (see eq.\eqref{hatFpropa1})  which respects the (axially extended)
diffeomorphisms (and, of course, can avoid the formidable obstacle of being
chiral). The desired handedness is obtained by taking the collapsing limit
$h_{\mu\nu}\to \frac {h_{\mu\nu}}2,f_{\mu\nu} \to \frac {h_{\mu\nu}}2$ (or
$h_{\mu\nu}\to \frac {h_{\mu\nu}}2,f_{\mu\nu} \to -\frac {h_{\mu\nu}}2$ for the
opposite handedness). This limit is smooth: we have not found any evidence of
singularity in it. This method admits different possible regularizations. We
have utilized two: the dimensional and the $\zeta$-function regularization, with
identical results. The  latter absolutely agree with the perturbative results
previously obtained in \cite{BGL,BDL,MAT1}.

On the basis of the evidence  collected so far, with no convincing
counterevidence, we conclude that not only does the parity odd trace anomaly
exist,
but all the procedures used in \cite{BGL,BDL,MAT1}
and the present paper are in accord\footnote{{We think the doubts
raised in
\cite{MAT1} in regard to the Pauli-Villars regularization, as being unable to
produce the same results, are worth a very detailed scrutiny.
Unfortunately, we are unable to say a final word on this issue due to the 
exceeding complexity of the calculation (at least in this particular case) 
and we have to postpone it to another occasion.}}.
It is reassuring in particular that there are
different ways of doing the same calculations while preserving chirality. 

{
Next let us comment on/recall some characteristics and possible consequences
of the odd trace anomaly.
Although we have done the calculation in 4d it is easy to see that a parity odd
trace anomaly may appear only in dimensions multiple of 4. Therefore, in
particular, they do not affect critical (super)string theories. Moreover, as was
already
pointed out in \cite{BGL}, the Pontryagin density vanish for 
a number of background metrics, among which the FRW one. But let us see the
possible consequences of the instances in which such anomaly does not vanish.} 
In this regards we cannot but repeat what was pointed out in the conclusion of
\cite{BGL}. The parity odd trace anomaly in Lorentzian metric has an imaginary
coefficient, which means in particular that the hamiltonian may be complex. This
may not be a problem as long as the fermion model is used in an effective field
theory context. A problem certainly
arises when gravity is itself quantized, because the lack of reality
(hermiticity) of the em tensor might propagate in the internal lines. Using this
anomaly as a selective criterion in the same way as chiral consistent gauge
anomalies were used in the
past, we should conclude that theories of massless Weyl fermions interacting
with gravity, with a definite unbalance of chiralities (an explicit example, the
old fashioned standard model, is shown on \cite{BGL}), should be excluded from
the realm of good theories, or at least very critically considered, because they
may turn out to be non-unitary\footnote{{A possibility might remain should
we consider PT invariance as the 
basic property, instead of hermiticity, see \cite{bender}, because the odd trace
anomaly is indeed PT invariant. However this requires a complete change of
paradigm for quantum field theory, which, to our best knowledge, has not yet
been
explored.}}. {Even though, as we just saw, critical (super)string theory is 
unaffected by the parity odd trace anomaly, any 4d theory which has is UV
completion in a
superstring theory should be completely anomaly free (and unitary) at any
intermediate 
energy regime from Planck all the way to low energy.} Finally, speaking of
unitarity, 
we cannot refrain from a
comment on a claim which is sometimes met in the literature: unitary theories
cannot have such kind of anomalies as the odd parity trace anomaly. Although we
believe the connection between unitary theories and absence of such anomalies is
true, we think the logical order should be reversed. One cannot impose unitarity
on a theory; unitarity must be the outcome of quantization. We think a more
sensible claim is: there are classical theories which are potentially unitary
(because they are based, say, on self-adjoint operators), but one has to verify
that unitarity persists after quantization; in this sense the absence of the
Pontryagin trace anomaly in a theory is a basic building block of its
unitarity. 

\vskip 1cm
{\bf Acknowledgements.}
L.B. would like to thank Fiorenzo Bastianelli and Claudio Corian\`o for a useful discussion.
This research has been supported 
by the University of
Rijeka under the research support No.~13.12.1.4.05
and in part by the Croatian
Science Foundation under the project No.~8946. The research of S.G. has been supported by the Israel Science Foundation (ISF), grant No. 244/17 in the last part of the project.

\vskip 1 cm

\appendix
\section*{Appendices}

\section{Bardeen's method}
\label{s:bardeen}

This appendix is a short account of Bardeen's method to derive gauge anomalies,
\cite{Bardeen}.

We consider a theory of Dirac fermions coupled to two non-Abelian (vector
$V_\mu$ and axial $A_\mu$) gauge potentials, both valued in a Lie algebra with
anti-hermitean generators $T^a$, with $[T^a,T^b]= f^{abc} T^c$. The action is
\be
S[V,A]= i\int d^4x \, \overline \psi \left( \slashed{\partial}+ \slashed{V} +
\gamma_5 \slashed {A}\right)\psi\label{SVA}
\ee
It is invariant under two sets of gauge transformations
\be
\left\{\begin{matrix} V_\mu \longrightarrow V_\mu + D_{V\mu} \alpha\\
        A_\mu \longrightarrow A_\mu+[A_\mu,\alpha]\\
\psi \longrightarrow (1-\alpha)\psi
       \end{matrix},\right. \quad\quad\quad
\left\{\begin{matrix} V_\mu \longrightarrow V_\mu+[A_\mu, \beta]\\
        A_\mu \longrightarrow A_\mu+D_{V\mu}\beta\\
\psi \longrightarrow (1+\gamma_5\beta)\psi
       \end{matrix}\right.\label{gaugetranf}
\ee
where $D_{V\mu}= \partial_\mu  +[V_\mu, ~\cdot~]$ and $\alpha =\alpha^a(x) T^a,
\beta=\beta^a(x) T^a$.

As a consequence there are two covariantly conserved currents, $j_\mu= j_\mu^a
T^a$ and
$ j_{5\mu}=j_{5\mu}^a T^a$, where
\be
j_\mu^a = \overline \psi \gamma_\mu T^a\psi,
\quad\quad j_{5\mu}^a = \overline \psi \gamma_\mu \gamma_5
T^a\psi\label{jmujmu5}
\ee
In the one-loop quantum theory it is impossible to preserve both conservations.
The most one
can do is to preserve, for instance, the vector one
\be
[D^\mu_V j_\mu]^a +[A^\mu, j_{5\mu}]^a=0\label{vectcons}
\ee
while the axial conservation becomes anomalous:
\be
[D^\mu_V j_{5\mu}]^a +[A^\mu, j_{\mu}]^a &=&
\frac 1{4\pi^2} \varepsilon_{\mu\nu\lambda\rho} \tr\left[T^a\left(
\frac 14 F_V^{\mu\nu} F_V^{\lambda\rho} +\frac 1 {12}  F_A^{\mu\nu}
F_A^{\lambda\rho}-
\frac 16  F_V^{\mu\nu}A^\lambda A^\rho \right.\right.\0\\
&&-\left.\left. \frac 16 A^\mu A^\nu F_V^{\lambda\rho} -\frac 23 A^\mu
 F_A^{\nu\lambda} A^\rho-\frac 13 A^\mu A^\nu A^\lambda
A^\rho\right)\right]\label{Bardeenanom}
\ee
where $F_V^{\mu\nu}= \partial^\mu V^\nu -\partial^\nu V^\mu+[V^\mu, V^\nu]+
{[A^\mu, A^\nu]}$, and
$  F_A^{\mu\nu}= \partial^\mu A^\nu -\partial^\nu A^\mu+[V^\mu, A^\nu]+ [A^\mu,
V^\nu]$.

From this expression we can derive two results in particular. Setting $A_\mu=0$
we
get the covariant anomaly
\be
[D^\mu_V j_{5\mu}]^a =\frac 1 {16\pi^2}  \varepsilon_{\mu\nu\lambda\rho}
 \tr\left(T^a F_V^{\mu\nu} F_V^{\lambda\rho}\right)\label{covanom}
\ee
Taking the collapsing limit $V\to \frac V2, A\to \frac V2$, and adding
\eqref{vectcons} to
\eqref{Bardeenanom}  we get
\be
[D_{V\mu} j^\mu_L]^a= \frac 1{24 \pi^2} \varepsilon_{\mu\nu\lambda\rho}
\tr\left[T^a\partial^\mu
\left(V^\nu\partial^\lambda V^\rho+\frac 12 V^\nu V^\lambda
V^\rho\right)\right]\label{consanom}
\ee
where $j_{L\mu}= \overline \psi_L \gamma_\mu\psi_L$, here $\psi_L = \frac
{1+\gamma_5}2 \psi$, which
is the consistent non-Abelian gauge anomaly.

\section{The axial-Riemannian geometry}

In this Appendix we collect the formulas, relevant to this paper, of
axial-Riemannian geometry. Such formulas have already appeared in \cite{MAT1},
although in a somewhat different notation.
An important difference with \cite{MAT1} is that, there, all the quantities
where functions of $x^\mu$. In this appendix, and throughout the paper they are
functions of $\hat x^\mu$ unless otherwise specified.

The main changes in notation are
\be
&&G_{\mu\nu}\longrightarrow \widehat g_{\mu\nu}, \quad\quad \hat
G^{\mu\nu}\longrightarrow \widehat g^{\mu\nu},\quad\quad \hat g \longrightarrow
\tilde g, \quad\quad \hat f \longrightarrow \tilde f\0\\
&& E_\mu^a \longrightarrow \widehat e_\mu^a,\quad\quad \hat E^\mu_a
\longrightarrow \widehat e^\mu_a,\quad\quad \hat e^\mu_a \longrightarrow  \tilde
e^\mu_a, \quad\quad \hat c^\mu_a \longrightarrow
\tilde c^\mu_a\0\\
&& \gamma_{\mu\nu}^\lambda \longrightarrow \Gamma_{\mu\nu}^\lambda, \quad\quad
\Gamma_{\mu\nu}^\lambda \longrightarrow \widehat \Gamma_{\mu\nu}^\lambda,
\quad\quad
\Omega_\mu^{ab} \longrightarrow \widehat \Omega_\mu^{ab},\quad\quad \Xi^\mu
\longrightarrow
\widehat \xi^\mu\0\\
&& {\cal R} \longrightarrow \widehat R,\quad\quad {\cal R}^{(1,2)}
\longrightarrow \widehat R^{(1,2)} \0
\ee

\subsection{Axial metric}

We use the symbols $g_{\mu\nu}, g^{\mu\nu}$ and $e_\mu^a, e_a^\mu$ in the usual
sense of metric and vierbein and their inverses, except for the fact that they
are functions of $\widehat x^\mu$. Then we introduce the MAT metric
\be
\widehat g_{\mu\nu}=g_{\mu\nu}+\gamma_5 f_{\mu\nu}\label{G}
\ee
where $f$ is a symmetric tensor.  {Their background values are $\eta_{\mu\nu}$
and
0, respectively. So, we write as usual $g_{\mu\nu}= \eta_{\mu\nu}+ h_{\mu\nu}$.}

In matrix notation the inverse of $\widehat g$, $\widehat g^{-1}$, is defined by
\be
\widehat g^{-1} = \tilde g +\gamma_5\tilde f,\quad\quad \widehat g^{-1} \widehat
g=1,\quad\quad
\widehat g^{\mu\lambda}\widehat g_{\lambda\nu}= \delta^\mu_\nu\label{GG-1}
\ee
which implies
\be
\tilde g f + \tilde f g=0, \quad\quad \tilde g g+\tilde f f =1.\label{GG-12}
\ee
So
\be
\tilde g=(1-g^{-1}\,f g^{-1}f)^{-1} g^{-1},\quad\quad  {\tilde f}=-(1-g^{-1}f
\,g^{-1}f)^{-1} g^{-1}f\, g^{-1}\label{G-12}
\ee

\subsection{MAT vierbein}

Likewise for the vierbein one writes
\be
\widehat e^a_\mu = e^a_\mu+ \gamma_5 c^a_\mu,\quad\quad \widehat e_a^\mu =\tilde
e_a^\mu+
\gamma_5\tilde c_a^\mu\label{vier1}
\ee
This implies
\be
\eta_{ab}\left(e^a_\mu e^b_\nu + c^a_\mu c^b_\nu\right)= g_{\mu\nu},
\quad\quad \eta_{ab}\left(e^a_\mu c^b_\nu + e^a_\nu c^b_\mu\right)=
f_{\mu\nu}\label{vier2}
\ee
Moreover, from $\widehat e_a^\mu \widehat e_\nu^a=\delta^\mu_\nu$,
\be
\tilde e_a^\mu c^a_\nu+\tilde c_a^\mu e^a_\nu =0,\quad\quad \tilde e_a^\mu
e^a_\nu+\tilde c_a^\mu c^a_\nu =\delta^\mu_\nu,\label{vier3}
\ee
one gets
\be
\tilde e_a^\mu = \left(\frac 1{1-e^{-1}c \,e^{-1} c}e^{-1}\right)_a^\mu
\label{vier4}
\ee
and
\be
\tilde c_a^\mu = -\left(e^{-1}c \frac 1{1-e^{-1}c \,e^{-1} c} e^{-1}c
e^{-1}\right)_a^\mu
\label{vier5}
\ee

\subsection{Christoffel and Riemann}

The ordinary Christoffel symbols are
\be
\Gamma_{\mu\nu}^\lambda= \frac 12 g^{\lambda\rho}\left( \partial_\mu
g_{\rho\nu}+\partial_\nu g_{\rho\mu} -\partial_\rho g_{\mu\nu}\right)
\label{chris}
\ee
The MAT Christoffel symbols are  defined in a similar way
\be
\widehat \Gamma_{\mu\nu}^\lambda&=& \frac 12 \widehat g^{\lambda\rho}\left(
\partial_\mu
\widehat g_{\rho\nu}+\partial_\nu \widehat g_{\rho\mu} -\partial_\rho \widehat
g_{\mu\nu}\right)
\label{Chris}\\
&=&\frac 12 \left( \tilde g^{\lambda\rho}\left( \partial_\mu
g_{\rho\nu}+\partial_\nu g_{\rho\mu} -\partial_\rho g_{\mu\nu}\right)+
 \tilde f^{\lambda\rho}\left( \partial_\mu f_{\rho\nu}+\partial_\nu f_{\rho\mu}
-\partial_\rho f_{\mu\nu}\right)\right)\0\\
&&+\frac 12 \gamma_5\left( \tilde g^{\lambda\rho}\left( \partial_\mu
f_{\rho\nu}+\partial_\nu f_{\rho\mu} -\partial_\rho f_{\mu\nu}\right)+
 \tilde f^{\lambda\rho}\left( \partial_\mu g_{\rho\nu}+\partial_\nu g_{\rho\mu}
-\partial_\rho g_{\mu\nu}\right)\right)\0\\
&\equiv& \Gamma^{(1)\lambda}_{\mu\nu} + \gamma_5 \Gamma^{(2)\lambda}_{\mu\nu}\0
\ee
where it is understood that $\partial_\mu=\frac {\partial}{\partial \hat
x^\mu}$, etc.

Proceeding the same way one can define the MAT Riemann tensor via $\widehat
R_{\mu\nu\lambda}{}^\rho $:
\be
\widehat R_{\mu\nu\lambda}{}^\rho &=& -\partial_\mu
\widehat\Gamma_{\nu\lambda}^\rho +
\partial_\nu \widehat\Gamma_{\mu\lambda}^\rho-\widehat \Gamma_{\mu\sigma}^\rho
\widehat\Gamma_{\nu\lambda}^\sigma +
\widehat\Gamma_{\nu\sigma}^\rho\widehat\Gamma_{\mu\lambda}^\sigma\label{Riem}\\
&=& -\partial_\mu \Gamma_{\nu\lambda}^{(1)\rho } + \partial_\nu
\Gamma_{\mu\lambda}^{(1)\rho }- \Gamma_{\mu\sigma}^{(1)\rho}
\Gamma_{\nu\lambda}^{(1)\sigma} +
\Gamma_{\nu\sigma}^{(1)\rho}\Gamma_{\mu\lambda}^{(1)\sigma}
-\Gamma_{\mu\sigma}^{(2)\rho} \Gamma_{\nu\lambda}^{(2)\sigma} +
\Gamma_{\nu\sigma}^{(2)\rho}\Gamma_{\mu\lambda}^{(2)\sigma}\0\\
&&+\gamma_5 \Big{(}  -\partial_\mu \Gamma_{\nu\lambda}^{(2)\rho } + \partial_\nu
\Gamma_{\mu\lambda}^{(2)\rho }- \Gamma_{\mu\sigma}^{(1)\rho}
\Gamma_{\nu\lambda}^{(2)\sigma} +
\Gamma_{\nu\sigma}^{(1)\rho}\Gamma_{\mu\lambda}^{(2)\sigma}
-\Gamma_{\mu\sigma}^{(2)\rho} \Gamma_{\nu\lambda}^{(1)\sigma} +
\Gamma_{\nu\sigma}^{(2)\rho}\Gamma_{\mu\lambda}^{(1)\sigma}\Big{)}\0\\
&\equiv& {\widehat R}^{(1)}_{\mu\nu\lambda}{}^\rho+\gamma_5 {\widehat
R}^{(2)}_{\mu\nu\lambda}{}^\rho\0
\ee
The MAT spin connection is introduced in analogy
\be
\widehat \Omega_\mu^{ab} &=& \widehat e_\nu^a\left(\partial_\mu\widehat e^{\nu
b}+\widehat e^{\sigma b}
\widehat \Gamma^\nu_{\sigma \mu}\right) = \Omega_\mu^{(1)ab}
+\gamma_5 \Omega_\mu^{(2)ab}\label{spinconn1}
\ee
where
\be
\Omega_\mu^{(1)ab}&=& e^a_\nu \left(\partial_\mu \tilde e^{\nu b}+ \tilde
e^{\sigma
b} \Gamma^{(1)\nu}_{\sigma \mu}+  \tilde c^{b\sigma}
\Gamma^{(2)\nu}_{\sigma \mu}\right)
+ c^a_\nu \left(\partial_\mu  \tilde  c^{\nu b}+ \tilde  e^{\sigma b}
\Gamma^{(2)\nu}_{\sigma \mu}+  \tilde c^{b\sigma}
\Gamma^{(1)\nu}_{\sigma \mu}\right)\label{Omega1}\\
\Omega_\mu^{(2)ab}&=& e^a_\nu \left(\partial_\mu  \tilde c^{\nu b}+ \tilde
e^{\sigma
b} \Gamma^{(2)\nu}_{\sigma \mu}+  \tilde  c^{b\sigma}
\Gamma^{(1)\nu}_{\sigma \mu}\right)
+ c^a_\nu \left(\partial_\mu   \tilde  e^{\nu b}+  \tilde  e^{\sigma b}
\Gamma^{(1)\nu}_{\sigma \mu}+ \tilde c^{b\sigma}
\Gamma^{(2)\nu}_{\sigma \mu}\right)\label{Omega2}
\ee

\subsection{Transformations. Diffeomorphisms}

We recall that under a diffeomorphism, $\delta x^\mu=\xi^\mu$, the ordinary
Christoffel symbols transform as tensors except for one non-covariant piece
\be
\delta_\xi^{(n.c.)}\Gamma_{\mu\nu}^\lambda=
\partial_\mu\partial_\nu\xi^\lambda\label{noncovgamma}
\ee

In the MAT context it is more opportune to introduces also axially-extended (AE)
diffeomorphisms. They are defined by
\be
\widehat x^\mu\rightarrow \widehat x^\mu+\widehat \xi^\mu(\widehat x^\mu),
\quad\quad\widehat
\xi^\mu=\xi^\mu+\gamma_5
\zeta^\mu\label{axialdiff}
\ee
Since operationally these transformations act in the same way as the usual
diffeomorphisms, it is easy to obtain
for the non-covariant part
\be
\delta^{(n.c.)}\widehat \Gamma_{\mu\nu}^\lambda=
\partial_\mu\partial_\nu\widehat \xi^\lambda\label{deltancGamma}
\ee
where the derivatives are understood with respect to $\widehat x^\mu$ and
$\widehat x^\nu$.
This means in particular that $\Gamma^{(2)\lambda}_{\mu\nu}$ is a tensor.

We have also
\be
\delta_{\widehat \xi}\widehat g_{\mu\nu} =\widehat D_\mu
\widehat\xi_\nu+\widehat
D_\nu\widehat \xi_\mu
\ee
where $\widehat\xi_\mu =\widehat g_{\mu\nu} \widehat\xi^\nu$ and ${\widehat
D}_\mu$ is the covariant derivative with respect to $\widehat \Gamma$.

In components one easily finds
\be
\delta_\xi g_{\mu\nu} &=&\xi^\lambda \partial_\lambda  g_{\mu\nu} + \partial_\mu
\xi^\lambda g_{\lambda\nu}+
\partial_\nu \xi^\lambda g_{\lambda\mu}\label{deltaxi}\\
\delta_\xi f_{\mu\nu} &=&\xi^\lambda \partial_\lambda  f_{\mu\nu} + \partial_\mu
\xi^\lambda f_{\lambda\nu}+
\partial_\nu \xi^\lambda f_{\lambda\mu}\0\\
\delta_\zeta g_{\mu\nu} &=&\zeta^\lambda \partial_\lambda  f_{\mu\nu} +
\partial_\mu \zeta^\lambda f_{\lambda\nu}+
\partial_\nu \zeta^\lambda f_{\lambda\mu}\label{deltazeta}\\
\delta_\zeta f_{\mu\nu} &=&\zeta^\lambda \partial_\lambda  g_{\mu\nu} +
\partial_\mu \zeta^\lambda g_{\lambda\nu}+
\partial_\nu \zeta^\lambda g_{\lambda\mu}\0
\ee

Summarizing
\be
&&\delta_\xi^{(n.c.)}\Gamma^{(1)\lambda}_{\mu\nu}=
\partial_\mu\partial_\nu\xi^\lambda,
\quad\quad\delta_\xi^{(n.c.)}\Gamma^{(2)\lambda}_{\mu\nu}=0\label{noncovGamma1}
\\
&&\delta_\zeta^{(n.c.)}\Gamma^{(1)\lambda}_{\mu\nu}=0, \quad\quad
\delta_\zeta^{(n.c.)}\Gamma^{(2)\lambda}_{\mu\nu} =
\partial_\mu\partial_\nu\zeta^\lambda\0
\ee
and the overall Riemann and Ricci tensors are tensor, and the Ricci
scalar ${ \widehat R}$ is a scalar. But also ${\widehat R}^{(1)}$ and ${\widehat
R}^{(2)}$,
separately, have the same tensorial properties.

\subsection{Transformations. Weyl transformations}

There are two types of Weyl transformations. The first is the obvious one
\be
\widehat g_{\mu\nu} \longrightarrow e^{2\omega}  \widehat g_{\mu\nu}, \quad\quad
\widehat g^{\mu\nu}
\to e^{-2\omega}\widehat g^{\mu\nu}\label{Weyl1G}
\ee
and
\be
\widehat  e_{\mu}^a \longrightarrow e^{\omega} \widehat e_{\mu}^a,
\quad\quad\widehat e^{\mu}_a \to
e^{-\omega}\widehat e^{\mu}_a\label{Weyl1E}
\ee
This leads to the usual relations
\be
\widehat\Gamma_{\mu\nu}^\lambda \longrightarrow \widehat\Gamma_{\mu\nu}^\lambda+
\partial_\mu
\omega\, \delta_\nu^\lambda + \partial_\nu \omega\, \delta_\mu^\lambda -
\partial_\rho \omega \,\widehat g^{\lambda\rho}\widehat g_{\mu\nu}
\label{Weyl1Gamma}
\ee
and
\be
\widehat\Omega_\mu^{ab} \longrightarrow\widehat\Omega_\mu^{ab} + \left(\widehat
e_\mu^a \widehat e^{\sigma
b} - \widehat e_\mu^b \widehat  e^{\sigma a} \right) \partial_\sigma \omega
\label{Weyl1Omega}
\ee

The second type of Weyl transformation is the axial one
\be
\widehat g_{\mu\nu} \longrightarrow e^{2\gamma_5 \eta}  \widehat g_{\mu\nu},
\quad\quad
\widehat g^{\mu\nu} \to e^{-2\gamma_5 \eta}\widehat g^{\mu\nu}\label{Weyl2G}
\ee
and
\be
\widehat e_{\mu}^a \longrightarrow e^{\gamma_5 \eta}  \widehat e_{\mu}^a,
\quad\quad
\widehat e^{\mu}_a \to e^{-\gamma_5 \eta}\widehat e^{\mu}_a\label{Weyl2E}
\ee
This leads to
\be
\widehat\Gamma_{\mu\nu}^\lambda \longrightarrow \widehat\Gamma_{\mu\nu}^\lambda+
\gamma_5\left(\partial_\mu \eta\, \delta_\nu^\lambda +
\partial_\nu \eta\, \delta_\mu^\lambda -
\partial_\rho \eta \,\widehat g^{\lambda\rho}\widehat g_{\mu\nu}\right)
\label{Weyl2Gamma}
\ee
and
\be
\widehat\Omega_\mu^{ab} \longrightarrow\widehat\Omega_\mu^{ab} + \gamma_5 \left(
\widehat e_\mu^a
\widehat e^{\sigma b} -\widehat e_\mu^b \widehat e^{\sigma a} \right)
\partial_\sigma \eta
\label{Weyl2Omega}
\ee
Eq.(\ref{Weyl2G}) implies
\be
g_{\mu\nu} \longrightarrow \cosh (2\eta) \, g_{\mu\nu} + \sinh(2\eta) \,
f_{\mu\nu},\quad\quad
f_{\mu\nu} \longrightarrow \cosh (2\eta) \, f_{\mu\nu} + \sinh(2\eta) \,
g_{\mu\nu}\label{getafeta}
\ee

We can write the axially-extended (AE) Weyl transformation in compact form using
the parameter $\widehat \omega=
\omega + \gamma_5 \eta$
 \be
\widehat g_{\mu\nu} \longrightarrow e^{2\widehat \omega}  \widehat g_{\mu\nu},
\label{axialWeyl}
\ee
etc.

\subsection{Volume density}

The ordinary density $\sqrt{g}$ is replaced by
\be
{\sqrt{\widehat g} =\sqrt{\det (\widehat g)} = \sqrt{\det (g+\gamma_5 f)}}
\label{volume1}
\ee
The expression in the RHS has to be understood as a formal Taylor expansion in
terms of the
axial-complex variable $g+\gamma_5 f$. This means
\be
 \tr\ln ( g+\gamma_5 f)
 &=& \frac {1+\gamma_5}2\,\tr \ln (g+f)  +\frac {1-\gamma_5}2\,  \tr\ln
(g-f)\label{volume2}
\ee
It follows that
\be
\sqrt{\widehat g}
= \frac 12 \left(\sqrt{\det(g+f)} +\sqrt{\det(g-f)}\right) + \frac {\gamma_5}2
   \left(\sqrt{\det(g+f)} -\sqrt{\det(g-f)}\right)
\label{volume3}
\ee
$\sqrt{\widehat g}$ has the basic property that, under AE diffeomorphisms,
\be
\delta_{\hat\xi} \sqrt{\widehat g}= \widehat\xi^\lambda \partial_\lambda
\sqrt{\widehat g} +
\sqrt{\widehat g}\,\partial_\lambda \widehat\xi^\lambda\label{volume4}
\ee
This is a volume density, and has the following properties
\be
\sqrt{\widehat g}\rightarrow e^{4\widehat \omega}\sqrt{\widehat
g},\label{volume5}
\ee
under an axial-Weyl transformations. Moreover
\be
\frac 1{\sqrt{\widehat g} }\partial_\nu  \sqrt{\widehat g} =\frac 12 \widehat
g^{\mu\lambda}
\partial_\nu\widehat
g_{\mu\lambda}=\widehat\Gamma_{\mu\nu}^\mu\label{Gammamumunu}
\ee

\section{Green's functions} 
\label{sec:prop}

In the text we have assumed the existence of the propagator $\widehat\EG$, the
inverse of 
$\widehat\EF$. In this Appendix we discuss this question by comparing it with
the ordinary case, as discussed in \cite{DeWitt1}. First we review the
approach of \cite{DeWitt1} in the ordinary gravity case. Then we explain the
modifications
required in the MAT case. We consider the case of a stationary metric and
axial-metric background.
We will assume eventually that the results hold also for nonstationary case,
provided the background varies mildly in time.

In this Appendix the flat gamma matrices are understood to be the Majorana ones,
that is, 
they are purely imaginary, together with $\gamma_5$: $\gamma_0\equiv \eta$ and
$\gamma_5$ 
are antisymmetric, while $\gamma_i$, $i=1,2,3$ are symmetric. 

\subsection{A summary of Green's functions}

Let us give first a short review of ordinary fermionic propagators, see
\cite{DeWitt1,DeWitt2,Christ1,Christ2}. 
We start from
\be
G(x,x') = \langle 0| {\cal T} \psi(x) \psi^\dagger (x')|0\rangle \label{Feyn} 
\ee 
This is not the standard Feynman Green function
\be
S_F(x,x') =  \langle 0| {\cal T} \psi(x) \bar \psi
(x')|0\rangle\label{stanFeyn} 
\ee
The two are related by $ S_F(x,x')=G(x,x') \eta$

Other Green functions 
are the advanced, $G^+(x,x')$,  and retarded,  $G^-(x,x')$; the positive and
negative frequency
Green functions, $G^{(+)}(x,x')$ and $G^{(-)}(x,x')$, respectively; and the
principal value Green function
$\bar G(x,x')= \frac 12 \left(G^+(x,x')+G^-(x,x')\right)$. The definitions
depends only on the contour of
integration of $p^0$ in the momentum space representation, while for the rest
they are the same. 
The important relation in this context is
\be
G(x,x') = \bar G(x,x')+ \frac i2 G^{(1)}(x,x'), \quad\quad  G^{(1)}= i\left
(G^{(+)}-G^{(-)}\right)\label{GG1}
\ee
For real fermions $\bar G(x,x')$ and $ G^{(1)}(x,x')$ are real. So they
represent the real and imaginary part of 
$ G(x,x')$. $ G^{(1)}(x,x')$ can be represented as
\be
 G^{(1)}(x,x')=  \langle 0| [ \psi(x), \psi^\dagger (x')]|0\rangle\equiv
\ES^{(1)} (x,x')\label{G1S1}
\ee
The Feynman propagator satisfies the equation
\be
i \sqrt{g} \eta\left( \gamma^\mu \nabla_\mu +m\right) G(x,x') = -{\bf
1}{\mathbf{\delta }}(x,x')\label{Fpropa}
\ee
and ${\bf 1}$ is the identity matrix in the spinor space. Both sides of
\eqref{Fpropa} transform as a bispinor density, i.e. 
like $ \sqrt{g} \gamma_0 \psi(x)$ at $x$ and as $\psi^\dagger(x')$ at $x'$.
Instead 
\be
i \sqrt{g} \eta \left( \gamma^\mu \nabla_\mu +m\right)
G^{(1)}(x,x')=0\label{Improp}
\ee
The approach of \cite{Christ1,Christ2} is based essentially on $G^{(1)}$.

Now let us make the ansatz
\be
G(x,x') = -i \left(  \gamma^\mu \nabla_\mu -m\right)\EG(x,x') \eta^{-1}
\label{ansatz1}
\ee
Inserting this into \eqref{Fpropa} one gets
\be
\sqrt{g} \left(\nabla_\mu g^{\mu\nu} \nabla_\nu- \left(m^2 +\frac 14 R\right)
\right) \EG(x,x')= - {\bf 1} \delta(x,x')  
\label{squareDirac}
\ee

Now we represent \eqref{squareDirac} as
\be
\int dx'' \EF(x,x'') \EG(x'',x') =  - {\bf 1} \delta(x,x')  \label{FG}
\ee
or, in operator form,
\be
\EF \, \EG =-1\label{FG1}
\ee
(understanding $\langle x| \EG|x' \rangle=\EG(x,x')$, etc.), where 
\be
\EF(x,x') = \sqrt{g} \left(\nabla_\mu g^{\mu\nu} \nabla_\nu- \left(m^2 +\frac
14 R\right)\right) {\bf 1} \delta(x,x') 
\label{EFx}
\ee
and the function and derivatives in the RHS are understood to be evaluated at
$x$.
Alternatively we represent \eqref{squareDirac} as
\be
\EF_x\, \EG(x,x') =  - {\bf 1} \delta(x,x')\label{FG2}
\ee
where $\EF_x$ is the differential operator acting on ${\bf 1} \delta(x,x') $ in
the RHS of \eqref{EFx}.

\subsection{Properties of $\EF$}

The operator $\EF$ in \eqref{squareDirac} is not selfadjoint. In fact
\be
\EF^\dagger = \gamma_0 \EF \gamma_0\label{EFdagger}
\ee
This implies that the construction of a Green's function is not straightforward.
In a stationary background a propagator is constructed out of modes which are
stationary eigenfunctions (plane waves, at least asymptotically)
with real frequencies. Given the Dirac equation 
\be
i(\gamma^\mu \nabla_\mu+m)u=0\label{Dirac+}
\ee
by suitably fixing the gauge for diffeomorphisms, one can always define a
complete set of eigenfunctions with real frequencies,
symbolically $u_+=\chi e^{-i\omega t}, u_- =\lambda e^{i\omega t}$, so that
(understanding the indices and integration over the space momenta)
\be
\psi = u_+a+ u_- a^\dagger\label{psiua}
\ee
where $a, a^\dagger$ are annihilation, creation operators (see chapter 19 of
\cite{DeWitt2}). 

In the same way one can infer the existence of an analogous complete set of
solutions, say $v_+,v_-$ of
\be
i(\gamma^\mu \nabla_\mu-m)v=0\label{Dirac-}
\ee

Now, even if $\EF$ is not self-adjoint, we can construct the following operator
\be
{\cal F}=\left(\begin{matrix} 0& \EF\\
          \EF^\dagger &0 
         \end{matrix}\right)\label{calF}
\ee
which is self-adjoint, and whose inverse is
\be
{\cal G}=\left(\begin{matrix} 0& \EG^\dagger\\
          \EG &0 
         \end{matrix}\right)\label{calG}
\ee
The mode solutions of ${\cal F}$ are
\be
\left(\begin{matrix} 0\\ u_+ \end{matrix}\right), \quad \left(\begin{matrix} 0\\
u_- \end{matrix}\right),\quad
\left(\begin{matrix}\gamma_0 v_+\\ 0 \end{matrix}\right), \quad
\left(\begin{matrix}\gamma_0 v_-\\ 0 \end{matrix}\right)
\label{modes}
\ee
which have all real frequencies. It follows that we can construct the Feynman
propagator of ${\cal F}$. Following
the argument of \cite{DeWitt2}, end of chapter 20, it has the form
\be
{\cal F}^{-1}= \left(\begin{matrix} 0& -\frac i{\EF^\dagger+i\epsilon}\\
         -\frac 1 {\EF+i\epsilon} &0 
         \end{matrix}\right)\label{calF-1}
\ee
Comparing with \eqref{calG} we get
\be
{\EG}=  -\frac 1{\EF+i\epsilon}\label{GF-1}
\ee

\subsection{Existence of mode functions}

The existence of mode functions, i.e. solutions of the Dirac equation
\eqref{Dirac+} of the type $u=\chi e^{i\omega t}$ with real $\omega$, in a
stationary background, is the basis for the existence of propagators. In
\cite{DeWitt2} the problem is discussed as follows. One shows that one can cast
\eqref{Dirac+} in the form
\be
Fu=0, \quad\quad F= \frac 12 \left\{ B^\mu, \frac {\partial}{\partial
x^\mu}\right\}- C\label{Fu}
\ee 
where
\be
B^\mu = i \eta \gamma^\mu, \quad\quad C= -\frac i4 \eta
 \{ \gamma^\mu, \omega_\mu\}\label{BandC}
\ee
The important thing is that, in the Majorana representation of the $\gamma$
matrices, 
$B^\mu$ is a symmetric matrix, while $C$ is
antisymmetric, and they are both purely imaginary.
By choosing the gauge $e_0^0=1,
e_0^i=0$ for the vierbein $e$, the operator $F$ becomes
\be
F=\frac 12 \left\{ B, \frac {\partial}{\partial t}\right\}- {\cal C}\label{Fugf}
\ee
where 
\be
B = i,\quad\quad {\cal C}= C -\frac 12 \left\{ B^i, \frac
{\partial}{\partial x^i}\right\}\label{BandCgf}
\ee
Again while $B$ is symmetric imaginary with $-iB$ being positive definite,
${\cal C}$ is antisymmetric
imaginary. Plugging the ansatz $u_A =\chi_A e^{-i\omega_A t}$ into
$Fu=0$ one gets the eigenvalue equation
\be
({\cal C} + i \omega_A B) \chi_A=0\label{eigenBC}
\ee
Due to the abovementioned propertis of $B$ and $\cal C$, one can find
eigenvalues and eigenvectors. The eigenvalues $\omega_A$ can be taken real and
positive.

\subsection{What changes when the background is MAT}

In this case the analog of \eqref{EFdagger} is
\be
{\widehat \EF}^\dagger =\eta\, \widehat \EF\,\eta
\label{FgFg'}
\ee
But as above we can proceed to construct the operator
\be
 {\widehat{\cal F}}=\left(\begin{matrix} 0& { \widehat\EF}\\
          {\widehat\EF}^\dagger &0 
         \end{matrix}\right)\label{hatcalF}
\ee
which is self-adjoint, and whose inverse is
\be
 {\widehat{\cal G}}=\left(\begin{matrix} 0&
 {\widehat\EG}^\dagger\\
          \overline{\widehat\EG} &0 
         \end{matrix}\right).\label{hatcalG}
\ee
Using the same argument as above we can conclude that
\be
 {\widehat{\EG}}=  -\frac
1{ {\widehat\EF}+i\epsilon}\label{hatGF-1}
\ee
The only delicate point in reaching this conclusion is the solutions of
\be
i\,{\widehat\gamma}^\mu {\widehat \nabla}_\mu \,u=0\label{hatDirac}
\ee
Eq.\eqref{Dirac+} is real, since  the gamma
matrices are purely imaginary. But, in
\eqref{hatDirac}, the presence of $\gamma_5$ poses a problem.  In a
representation in which the gamma matrices are purely imaginary, the $\gamma_5$
is also imaginary, thus eq.\eqref{hatDirac} is complex, and, based on the
analogy with the previous subsection, one cannot be sure a priori
that there are real frequency solutions. However we notice that the operator
$\eta \widehat F$ is self-adjoint. This remark lends us a way out.

Another crucial point is the gauge fixing, so that one can end up with something
analog to
\eqref{BandCgf}, in which 
$-iB$ is positive definite. As we saw above, this is obtained by choosing in
particular $e_0^0=1,
e_0^i=0$.
In MAT the coefficient of $\gamma^0$ is $\widehat e_0^\mu$, which contains also
$\gamma_5 c_0^\mu$.
We shall choose $c_0^\mu=0$. As a consequence the analog of $Fu=0$ is $\widehat
F \widehat u=0$ where 
\be
\hat F=\frac 12 \left\{ \widehat B, \frac {\partial}{\partial t}\right\}-
\widehat{\cal C}\label{Fugfhat}
\ee
where $\widehat B=B$, i.e. symmetric and such that $-iB$ is positive definite.
As for $\widehat{\cal C}$, it can be written as
\be 
\widehat{\cal C}= \widehat{\cal C}_a + \widehat{\cal C}_s\label{calC}
\ee
where $\widehat{\cal C}_a$ is imaginary  antisymmetric and does not contain
$\gamma_5$, while 
$ \widehat{\cal C}_s$ is real, linear in $\gamma_5$ and symmetric. However
altogether it is self-adjoint. 

Plugging the ansatz $\widehat u_A =\widehat \chi_A e^{-i\omega_A t}$ into
$\eta {\widehat F}\widehat u=0$ one gets the equation
\be
(\widehat {\cal C} - \omega_A ) \widehat \chi_A=0\label{eigenCA}
\ee
which is an eigenvalue equation for $\widehat {\cal C} $. Since the latter is
self-adjoint we know there exists a complete set of eigenfunctions. This is what
we need. 

So the remaining question is: is the choice  $c_0^\mu=0$ permitted? 
In order to see this one has to check that the defining equations
(\ref{vier1},\ref{vier2}) for the axial-complex vierbein and the like in Appendix B are still
valid. Now, suppose the ordinary gauge fixed vierbein satisfies such defining
equation (which they do
in \cite{DeWitt1}).
Then we can set the axial-imaginary vierbein $c$ and $c^{-1}$ to 0, while
preserving the defining relations.
In other words, there is a large gauge freedom, and in particular we can choose 
$c_0^\mu=0$.



\begin{thebibliography}{99}

\bibitem{MAT1}
  L.~Bonora, M.~Cvitan, P.~Dominis Prester, A.~Duarte Pereira, S.~Giaccari and
T.~\v{S}temberga,
{\it Axial gravity, massless fermions and trace anomalies,}
  Eur.\ Phys.\ J.\ C {\bf 77} (2017) no.8, 511
  [arXiv:1703.10473 [hep-th]].

\bibitem{BGL}
  L.~Bonora, S.~Giaccari and B.~Lima de Souza,
  {\it Trace anomalies in chiral theories revisited,}
  JHEP {\bf 1407}, 117 (2014)
  [arXiv:1403.2606 [hep-th]].	

\bibitem{BDL}  L.~Bonora, A.~D.~Pereira and B.~L.~de Souza,
 {\it Regularization of energy-momentum tensor correlators and parity-odd
terms,}
  JHEP {\bf 1506}, 024 (2015)
  [arXiv:1503.03326 [hep-th]].

\bibitem{Bardeen} W.A. Bardeen, {\it Anomalous Ward
Identities in Spinor Field Theories}, Phys.\ Rev.\ {\bf 184}  1848 (1969).

\bibitem{DeWitt1} B.~S.~DeWitt, {\it Dynamical theory of groups and fields},
Gordon and Breach, New York, 1965.

\bibitem{DeWitt2} B.~S.~DeWitt, {\it Global approach to quantum field theory},
vol.I and II.
 
\bibitem{capper} D.~M.~Capper and M.~J.~Duff, {\it Trace Anomalies in
Dimensional Regularization}, Nuovo Cim.\ {\bf 23 A} (1974) 173; {\it Conformal
anomalies and the renormalizability problem in quantum gravity}, Phys.\ Lett.\
{\bf 53 A} (1975) 361.

\bibitem{deser} S.~Deser, M.~J.~Duff and C.~J.~Isham, {\it Non-local conformal
anomalies},
Nucl.\ Phys.\ B {\bf 111} (1976) 45.

\bibitem{bernard} C.~Bernard and A.~Duncan, {\it Regularization and
renormalization of quantum field theory in curved space-time}, Ann.\ Phys.\ {\bf
107} (1977) 201.

\bibitem{brown} L.~S.~Brown, {\it Stress-tensor trace anomaly ina a
gravitational metric: scalar fields}, Phys.\ Rev.\ {\bf D 15} (1977) 1469.

\bibitem{brown-cas} L.~S.~Brown and J.~P. Cassidy, {\it Stress-tensor trace
anomaly in a gravitational metric: General theory, Maxwell field}, Phys.\ Rev.\
{\bf D 15} (1977) 2810.
{\it Stress tensors and their trace anomalies in conformally flat space-time},
Phys.\ Rev.\ {\bf D 16} (1977) 1712.

\bibitem{Christ1}  S.~M.~Christensen, {\it Vacuum expectation value of the
stress tensor in an arbitrary curved background: The covariant point-separation
method}, Phys.\ Rev.\ {\bf D 14} (1976) 2490.
 
\bibitem{Christ2} S.~M.~Christensen {\it Regularization, renormalization and
covariant geodesic point separation},
Phys.\ Rev.\ {\bf D17} (1978) 946.

\bibitem{adler} S.~ L.~ Adler, J.~ Lieberman, Y.~J.~ Ng, {\it Regularization of
the stress-energy tensor for vector and scalar particles propagating in a
general background metric}, Ann.\ Phys.\ {\bf 106} (1977) 209.

 
\bibitem{duff1} M.~J.~Duff, {\it Observations on conformal anomalies}, Nucl.\
Phys.\
{\bf B125} (1977) 334.

\bibitem{dowker} J.~S.~Dowker and R.~ Critchley. {\it Stress-tensor conformal
anomaly for scalar, spinor, and vector fields},  Phys.\ Rev.\ {\bf D 16} (1977)
3390.

\bibitem{tsao} H.~-S.~Tsao, {\it Conformal anomaly in a general background
metric}, Phys.\ Lett.\ {\bf 68B} (1977) 79.

\bibitem{ChD} S.~M.~Christensen and M.~J.~Duff {\it Axial and conformal
anomalies
for arbitrary spin in gravity and supergravity}, Phys.\ Lett.\ {\bf 76B} (1978)
571.

\bibitem{vilenkin} A.~Vilenkin, {\it Pauli-Villars Regularization and Trace
Anomalies}, 
Nuovo Cim.\ {\bf A44} (1978) 441. 	
  
\bibitem{wald} R.~M.~Wald, {\it Axiomatic renormalization of the stress tensor
of a conformally invariant field in conformally flat spacetimes}, Ann.\ Phys.\
{\bf 110} (1978) 472. 
{\it Trace anomaly of a conformally invariant quantum field in curved
spacetime}, Phys.\ Rev.\ {\bf D 17} (1978) 1477.
   
\bibitem{CD2} S.~M.~Christensen and M.~J.~Duff, 
{\it New gravitational index theorems and super theorems},
Nucl.\ Phys.\ {\bf B154} (1979) 301 

\bibitem{BCR} L.~Bonora, P.~ Cotta-Ramusino and C.~Reina, {\it Conformal anomaly
and cohomology},
Phys.\ Lett.\ {\bf 126B} (1983), 305 

\bibitem{BPT1} L.~Bonora, P.~Pasti, M.~Tonin,
{\it Gravitational and Weyl anomalies}, Phys.\ Lett.\ {\bf 149B} (1985), 346.

\bibitem{BPT2} L.~Bonora, P.~Pasti, M.~Tonin,
{\it The anomaly structure of theories with external gravity}, J.\ Math.\ Phys.\
{\bf
27} (1986), 2259

\bibitem{BBP} L.~Bonora, M.~Bregola, P.~Pasti, {\it Weyl cocycles},
Class.\ Quant.\ Grav.\ {\bf 3} (1986) 635.

\bibitem{Osborn1}
 H.~Osborn , {\it Weyl consistency conditions and a local renormalization group
equation 
for general renormalizable field theoriees}, Nucl.\ Phys.\ {\bf B363} (1991)
486.

\bibitem{Osborn2}
  H.~Osborn and A.~C.~Petkos,
{\it Implications of conformal invariance in field theories for general
dimensions,}
  Annals\ Phys.\  {\bf 231} (1994) 311
  [hep-th/9307010].

\bibitem{Duff93}
  M.~J.~Duff,
{\it Twenty years of the Weyl anomaly,}
  Class.\ Quant.\ Grav.\  {\bf 11} (1994) 1387
  [hep-th/9308075].

\bibitem{Deser}
  S.~Deser and A.~Schwimmer,
{\it Geometric classification of conformal anomalies in arbitrary dimensions,}
  Phys.\ Lett.\ B {\bf 309} (1993) 279
  [hep-th/9302047].

\bibitem{Bzowski}
  A.~Bzowski, P.~McFadden and K.~Skenderis,
{\it Renormalised 3-point functions of stress tensors and conserved currents in
CFT,}
  arXiv:1711.09105 [hep-th].

\bibitem{Godazgar}
  H.~Godazgar and H.~Nicolai,
{\it A rederivation of the conformal anomaly for spin-1/2,}
  Class.\ Quant.\ Grav.\  {\bf 35} (2018) no.10,  105013
  [arXiv:1801.01728 [hep-th]].

\bibitem{Coriano}
  C.~Corianò and M.~M.~Maglio,
{\it Renormalization, Conformal Ward Identities and the Origin of a Conformal
Anomaly Pole,}
  Phys.\ Lett.\ B {\bf 781} (2018) 283
  [arXiv:1802.01501 [hep-th]].

\bibitem{BL} L.~Bonora and B.~L.~de Souza,
  {\it Pure contact term correlators in CFT,} Proc. 18th Bled Workshop "What
Comes Beyond Standard Models", Bled 2015.  [arXiv:1511.06635 [hep-th]].

\bibitem{Bast} F.~Bastianelli and R.~Martelli,
  {\it On the trace anomaly of a Weyl fermion,}
  JHEP {\bf 1611} (2016) 178
  [arXiv:1610.02304 [hep-th]].

\bibitem{Shapiro} S.~Mauro and I.~L.~Shapiro,
  {\it Anomaly-induced effective action and Chern-Simons modification of
general
relativity,}
  Phys.\ Lett.\ B {\bf 746} (2015) 372
  [arXiv:1412.5002 [gr-qc]].

\bibitem{Nakayama1}
Y.~Nakayama, {\it CP-violating CFT and trace anomaly,}
Nucl.\ Phys.\ B {\bf 859} (2012) 288.

\bibitem{Nakayama2}
  Y.~Nakayama,
{\it On the realization of impossible anomalies,}
  arXiv:1804.02940 [hep-th].	

\bibitem{Bertlmann} R.~A. ~Bertlmann, {\it Anomalies in quantum field theory},
Oxford Science Publications, Oxford
U.K. (1996).

\bibitem{Fujikawa} K.~ Fujikawa and H.~ Suzuki, {\it Path integrals and quantum
anomalies}, Oxford Science
Publications, Oxford U.K. (2004).

\bibitem{BastVan} F.~ Bastianelli and P. ~Van Nieuwenhuizen, {\it Path integrals
and anomalies in curved space},
Cambridge University Press, Cambridge U.K. (2009).

\bibitem{AB1} A.~Andrianov, L.~Bonora,
{\it Finite-mode regularization of the fermion functional integral.I},
Nucl.\ Phys.\ {\bf B233} (1984), 232.

\bibitem{AB2} A.~Andrianov, L.~Bonora,
{\it Finite-mode regularization of the fermion functional integral.II},
Nucl.\ Phys.\ {\bf B233} (1984), 247

\bibitem{Grabowska}  D.~M.~Grabowska and D.~B.~Kaplan,
  {\it Chiral solution to the Ginsparg-Wilson equation,}
  Phys.\ Rev.\ D {\bf 94} (2016) no.11,  114504
  [arXiv:1610.02151 [hep-lat]].

\bibitem{Hess-Greiner}
P.~O.~Hess and W.~Greiner,
{\it Pseudo-complex General Relativity,}
  Int.\ J.\ Mod.\ Phys.\ E {\bf 18}, 51 (2009)
  [arXiv:0812.1738 [gr-qc]].

P.~O.~Hess and W.~Greiner,
 {\it Pseudo-Complex Field Theory,}
  Int.\ J.\ Mod.\ Phys.\ E {\bf 16}, 1643 (2007)
  [arXiv:0705.1233 [hep-th]].

\bibitem{Crumey} A~. Crumeyrolle, {\it Variétés différentiables 
à coordonnées hypercomplexes. Application à une géométrisation et à 
une généralisation de la théorie d'Einstein-Schrodinger.}
Ann.\ de la Fac.\ des Sciences de Toulouse, $4^e$ s\'erie, {\bf 26} (1962) 105.

A~. Crumeyrolle, {\it Sur quelques interpr\'etations physiques et th\'eoriques
des equations 
du champ unitaire d'Einstein-Sch${\rm {\ddot o}}$dinger,}
Riv.\ Mat.\ Univ.\ Parma (2) {\bf 5} (1964) 85.

\bibitem{Clerc} R.~-L.~ Clerc {\it Résolution des équations aux connexions du
cas
antisymétrique de la théorie unitaire hypercomplexe.
Application à un principe variationnel},
Ann.\ de l'I.H.P.\ Section A, {\bf 12} no. 4, (1970) 343.

R.~-L.~ Clerc {\it  Équations de champ symétriques et équations du
mouvement sur une variété pseudo-riemannienne à connexion non symétrique},
Ann.\ de l'I.H.P.\ Section A, {\bf 17} no. 3, (1972) 227.

\bibitem{Mantz}
  C.~Mantz and T.~Prokopec,
{\it Hermitian Gravity and Cosmology,}
  arXiv:0804.0213 [gr-qc].

\bibitem{bender}  C.~M.~Bender,
{\it Making sense of non-Hermitian Hamiltonians,}
  Rept.\ Prog.\ Phys.\  {\bf 70} (2007) 947
    [hep-th/0703096 [HEP-TH]].


\end{thebibliography}
\end{document}